\documentclass[useAMS,usenatbib,aas_macros,fleqn]{mn2e}
\usepackage{amsmath}
\usepackage{graphicx} 
\usepackage{cleveref}
\usepackage{subfig}
\usepackage{bm}
\usepackage{breqn}

\usepackage{color}

\newcommand{\equ}[1]{eq.~(\ref{eq:#1})}

\newcommand{\se}[1]{\S\ref{sec:#1}}
\newcommand{\fig}[1]{Fig.~\ref{fig:#1}}
\newcommand{\figs}[1]{Figs.~\ref{fig:#1}}
\newcommand{\Fig}[1]{Figure~\ref{fig:#1}}

\newcommand{\be}{\begin{equation}}
\newcommand{\ee}{\end{equation}}
\newcommand{\ba}{\begin{align}}
\newcommand{\ea}{\end{align}}
\newcommand{\bad}{\begin{equation} \begin{aligned}}
\newcommand{\ead}{\end{aligned} \end{equation}}
\newcommand{\bea}{\begin{eqnarray}}
\newcommand{\eea}{\end{eqnarray}}

\newcommand{\Msun}{M_\odot}

\newcommand{\ifm}[1]{\relax\ifmmode#1\else$\mathsurround=0pt #1$\fi}
\newcommand{\kms}{\ifmmode\,{\rm km}\,{\rm s}^{-1}\else km$\,$s$^{-1}$\fi}

\newcommand{\kpc}{\,{\rm kpc}}

\newcommand{\ltsima}{$\; \buildrel < \over \sim \;$}
\newcommand{\lsim}{\lower.5ex\hbox{\ltsima}}
\newcommand{\gtsima}{$\; \buildrel > \over \sim \;$}
\newcommand{\gsim}{\lower.5ex\hbox{\gtsima}}

\def\omm{\Omega_{\rm m}}
\def\oml{\Omega_{\Lambda}}
\def\omb{\Omega_{\rm b}}

\def\Mv{M_{\rm vir}}

\def\Rv{R_{\rm v}}
\def\rs{r_{\rm s}}
\def\Vv{V_{\rm v}}
\def\Mstar{M_\star}
\def\Ms{M_\star}
\def\Mg{M_{\rm gas}}
\def\re{r_{\rm 1/2}}

\def\rp{r_{\rm p}}
\def\Vp{V_{\rm p}}
\def\tdyn{t_{\rm dyn}}
\def\mubar{ \tilde{\mu} }
\def\rhobar{ \bar{\rho} }
\def\thetam{ \theta_{\rm m} }

\def\DE{\Delta E}

\def\Eb{E_{\rm b}}
\def\Uin{U_{\rm in}}
\def\Uout{U_{\rm out}}
\def\mv{m_{\rm v}}
\def\Dm{\Delta m}
\def\lv{l_{\rm v}}

\def\lE{l_{\rm E}}
\def\lt{l_{\rm t}}
\def\ltr{l_{\rm tr}}

\def\le{l_{\rm e}}
\def\lrp{l_{\rm rp}}
\def\vc{v_{\rm c}}
\def\vv{v_{\rm v}}
\def\sigmar{\sigma_{\rm r}}
\def\sigmat{\sigma_{\rm t}}
\def\mpeak{m_{\rm peak}}
\def\msub{m_{\rm sub}}
\def\mres{m_{\rm res}}
\def\mstar{m_\star}
\def\mg{m_{\rm gas}}
\def\xc{x_{\rm c}}
\def\rc{r_{\rm c}}
\def\jc{j_{\rm c}}
\def\zpeak{z_{\rm peak}}
\def\zroot{z_{\rm root}}

\def\nS{n_{\rm Sersic}}

\def\fgs{f_{\rm gas,sat}}
\def\fgh{f_{\rm gas,host}}
\def\cNFW{c_{\rm NFW}}
\def\aE{{\alpha_{\rm Ein}}}
\def\cD{c_{\rm D}}
\def\aD{{\alpha_{\rm D}}}
\def\lamh{\lambda_{\rm halo}}
\def\rmd{{\rm d}}

\newcommand{\apjl}{ApJL}
\newcommand{\apjs}{ApJS}

\newcommand{\aap}{A\&A}

\title[Formation of UDGs]
{Formation of ultra-diffuse galaxies in the field and in galaxy groups}

\author[F. Jiang et al.]
{\parbox[t]{\textwidth}
{ 
Fangzhou Jiang$^{1}$\thanks{E-mail: fangzhou.jiang@mail.huji.ac.il},
Avishai Dekel$^{1,2}$\thanks{E-mail: avishai.dekel@mail.huji.ac.il},
Jonathan Freundlich$^{1}$,\\
Aaron J. Romanowsky$^{3,4}$,
Aaron Dutton$^{5}$,
Andrea Macci\`o$^{5,6}$,
Arianna Di Cintio$^{7,8}$
}
\\ \\ 
$^1$ Centre for Astrophysics and Planetary Science, Racah Institute of Physics,
The Hebrew University, Jerusalem 91904, Israel\\
$^2$ SCIPP, University of California, Santa Cruz, CA 95064,
USA\\
$^3$ Department of Physics \& Astronomy, San Jos\'e State University, One Washington Square, San Jose, CA 95192, USA\\
$^4$ University of California Observatories, 1156 High Street, Santa Cruz, CA 95064, USA\\
$^5$ New York University Abu Dhabi, PO Box 129188, Abu Dhabi, United Arab Emirates\\
$^6$ Max Planck Institute f${\ddot{u}}$r Astronomie, K${\ddot{o}}$nigstuhl 17, D-69117 Heidelberg, Germany\\
$^{7}$ Instituto de Astrof\'{i}sica de Canarias, Calle Via L\'{a}ctea s/n, E-38206 La Laguna, Tenerife, Spain\\
$^{8}$ Universidad de La Laguna. Avda. Astrof\'{i}sico Fco. S\'{a}nchez, La Laguna, Tenerife, Spain
}

\begin{document}

\large

\pagerange{\pageref{firstpage}--\pageref{lastpage}} \pubyear{2018}

\maketitle

\label{firstpage}

\begin{abstract}
We study ultra-diffuse galaxies (UDGs) in zoom in cosmological simulations, seeking the origin of UDGs in the field versus galaxy groups. 
We find that while field UDGs arise from dwarfs in a characteristic mass range by multiple episodes of supernova feedback (\citealt{dicintio17}), group UDGs may also form by tidal puffing up and they become quiescent by ram-pressure stripping.
The field and group UDGs share similar properties, independent of distance from the group centre. 
Their dark-matter haloes have ordinary spin parameters and centrally dominant dark-matter cores. 
Their stellar components tend to have a prolate shape with a S\'ersic index $n\sim 1$ but no significant rotation.
Ram pressure removes the gas from the group UDGs when they are at pericentre, quenching star formation in them and making them redder. 
This generates a colour/star-formation-rate gradient with distance from the centre of the dense environment, as observed in clusters.
We find that $\sim20$ per cent of the field UDGs that fall into a massive halo survive as satellite UDGs. 
In addition, normal field dwarfs on highly eccentric orbits can become UDGs near pericentre due to tidal puffing up, contributing about half of the group-UDG population.
We interpret our findings using simple toy models, showing that gas stripping is mostly due to ram pressure rather than tides.
We estimate that the energy deposited by tides in the bound component of a satellite over one orbit can cause significant puffing up provided that the orbit is sufficiently eccentric. 
\end{abstract}


\section{Introduction}
\label{sec:intro}

\smallskip
Ultra-diffuse galaxies (UDGs) are low-surface brightness systems ($\mu_{0}(g)>$24 mag arcsec$^{-2}$) with surprisingly large effective radii ($\re>1.5$kpc).
They have stellar masses similar to those of dwarf galaxies and their surface density profiles show similar S\'ersic indices to those of disk galaxies (e.g., \citealt{mowla17}, \citealt{greco18}). 
UDGs are ubiquitous in clusters and groups (e.g., \citealt{vd15a}; \citealt{merritt16}; \citealt{koda15}; \citealt{yagi16}; \citealt{janssens17}), but they are also found in the field (e.g., \citealt{md16}; \citealt{rt17}; \citealt{leisman17}). 
In dense environments, UDGs exhibit intermediate-to-old stellar populations, based on spectroscopic studies of a few cases \citep{fm18,gu18,rl18}.
In the field, UDGs seem to show younger stellar populations \citep{pandya18}, evidence of ongoing star formation, irregular morphologies, and high gas fractions that are typical of dwarf galaxies in the field (\citealt{shi17,greco18}; but see also \citealt{papastergis17}).  

\smallskip
There is still no consensus yet regarding the host-halo mass of UDGs.
In a couple of case studies, the inferred host halo mass is comparable to that of the Milky-Way halo \citep{beasley16,vd16}.
However, these estimates are liable to the applicability of the empirical dynamical mass estimator \citep{wolf10} on UDGs, and also to the extrapolation of the halo mass profile from the location of the kinematic tracers to the virial radius. 
More robust evidence of the high halo mass of the few UDGs lies in the fact that they have higher abundance of globular clusters (GCs) than dwarf galaxies of similar stellar mass (e.g., \citealt{vd17}), and that the abundance of GCs are known to scale tightly with halo mass \citep{harris17}.
However, UDGs show a large variance in their GC-abundance \citep{lim18,amorisco18gc}, and the ubiquity of UDGs in the Coma cluster imply that UDGs cannot all dwell in Milky-Way-mass haloes \citep{amorisco18mass}. 

\smallskip
In addition to the halo-mass debate, the more general open question is: are UDGs distinctive in any parameter space compared to ``normal'' galaxies, or are they simply the tails of unimodal distributions?
For example, it is not clear yet whether or not UDGs form a distinct mode in the $\re$ distribution of all galaxies in the mass range of dwarf galaxies (but see \citealt{danieli18}); and, related, regarding the standard picture that galaxy size is proportional to host halo spin and virial radius, whether or not UDGs constitute the high-spin tail of dwarf galaxies \citep{al16,rong17}.
Moreover, UDGs have low S\'ersic indices, raising the question whether they are simply the faint end of oblate galaxies or not. 

\smallskip
Theoretical studies of UDGs are quite preliminary. 
The main challenge lies in generating a statistical sample of UDGs in cosmological simulations.
If UDGs are dwarf-sized objects in terms of halo mass (i.e., $\sim10^{10}\Msun$), resolving a UDG as a satellite in a Coma-sized host is computationally expensive: it requires a dynamical range in mass of more than 5 orders of magnitudes. 
Producing a statistical sample of field dwarfs is easier.
\cite{dicintio17} first identified UDGs in $\Lambda$CDM simulations \citep{wang15}, and showed that their host haloes are in a narrow mass range of $\Mv=10^{10-11}\Msun$, and implied that the formation of field UDGs are associated with episodic supernovae (SNe) outflows.
\cite{chan18} manually strangulated the gas supply of simulated field galaxies in order to mimic what happens to satellite galaxies in a dense environment, and found that SNe outflows before the `strangulation' together with the passively aging stellar population can give rise to red UDGs, depending on when the quenching is imposed. 
This approach is still not a full-fledged treatment of a dense environment, neglecting the details of tidal effects and ram-pressure stripping, and is limited to a small sample. 
Semi-analytic models can generate galaxies that satisfy the size and surface-brighness criteria of UDGs (e.g., \citealt{al16}, \citealt{rong17}), but such galaxies lie almost exclusively on the high-halo-spin tail of dwarf galaxies, and largely reflect the input of the semi-analytic recipe for galaxy size. 

\smallskip
In this paper, we study both field UDGs and satellite UDGs in zoom-in cosmological simulations.
We elaborate on the simulation suite used by \cite{dicintio17}, and present new findings about the shape, stellar population, and host halo structure of the field UDGs. 
We also identify UDGs as satellites in a zoom-in simulation of a galaxy group \citep{dutton15}, characterize the properties of UDGs as a function of group-centric distance, and explore their formation mechanisms. 

\smallskip
The outline of this paper is as follows.
In \se{method} we describe the simulations and how we perform the measurements.
In \se{field} and \se{group}, we present the results for field UDGs and group UDGs, respectively.
In \se{discussion}, we use analytic toy models to clarify UDG-formation mechanisms inferred from the simulation results.
In \se{conclusion}, we summarize our findings.

\section{Method}
\label{sec:method}

\subsection{Simulations}
\label{sec:sim}

\smallskip
Our sample of field galaxies is taken from the NIHAO suite \citep{wang15}, consisting of 90 galaxies with halo mass in the range of $\Mv(z=0) = 10^{9.5-12.3} \Msun$ that are evolved using the SPH code {\tt Gasoline 2.0} \citep{wadsley17}.
The code includes subgrid prescriptions for turbulent mixing of metals and energy \citep{wadsley08}, and cooling via emission lines in a uniform ultraviolet ionizing background as described in \cite{shen10}. 
Star formation and feedback follow the recipe used in the MaGICC simulations \citep{stinson13}: gas is eligible to form stars according to the Kennicutt-Schmidt Law when its density is higher than $n_{\rm th} = 10.3$ cm$^{-3}$ and temperature lower than $15000$K; stars feed energy back into the interstellar medium via blast-wave SNe feedback \citep{stinson06} and pre-SNe stellar feedback from massive stars. 
The simulations are run in a flat $\Lambda$CDM cosmology with parameters from the Planck Collaboration (\citealt{planck15};
 $\omm = 0.3175$, $\oml = 0.6824$, $\omb  = 0.0490$, $h=0.671$, $\sigma_8 = 0.8344$, $n = 0.9624$).
Force softening and particle mass depend on the mass of the galaxy, chosen such that the density profile is well resolved down to 1 per cent of the virial radius. 
For the UDGs in the NIHAO simulations, the typical value of force softening is 132.6 pc, and typical particle mass for gas is $1.173\times10^4\Msun$.

\smallskip
Our sample of satellite galaxies are taken from a system of virial mass $\Mv(z=0)=10^{13.33}\Msun$ that was originally used by \cite{dutton15} to study the host halo response to bright central galaxies (`halo4.2' in \citealt{dutton15}).
The star formation and feedback prescriptions are similar to those used in the NIHAO simulations, except that the star formation threshold is calculated as $n_{\rm th} = 32(m_{\rm gas}/5)/\epsilon^3 =1.16$ cm$^{-3}$, where $m_{\rm gas}=10^6\Msun$ is the initial gas particle mass; $(m_{\rm gas}/5)$ is the minimum gas particle mass; and $\epsilon=606$ comoving pc is the spatial resolution.
A typical UDG has $\mstar\sim10^8\Msun$ and $\re\sim 3$kpc \citep{vd15a} -- if such systems exist in the simulation, they would be adequately resolved with more than $100$ star particles and with their effective radii equal to $\sim 5(1+z)$ times $\epsilon$.  
The simulation is run in a flat $\Lambda$CDM cosmology with parameters from the Wilkinson Microwave Anisotropy Probe 7th year \citep{wmap7} results  ($\omm = 0.2748$, $\oml = 0.7252$, $\omb  = 0.0458$, $h=0.702$, $\sigma_8 = 0.816$, $n = 0.968$). 
Feedback from AGN is not included. 

\subsection{Analysis}
\label{sec:analysis}

\begin{figure}
\includegraphics[width=0.5\textwidth]{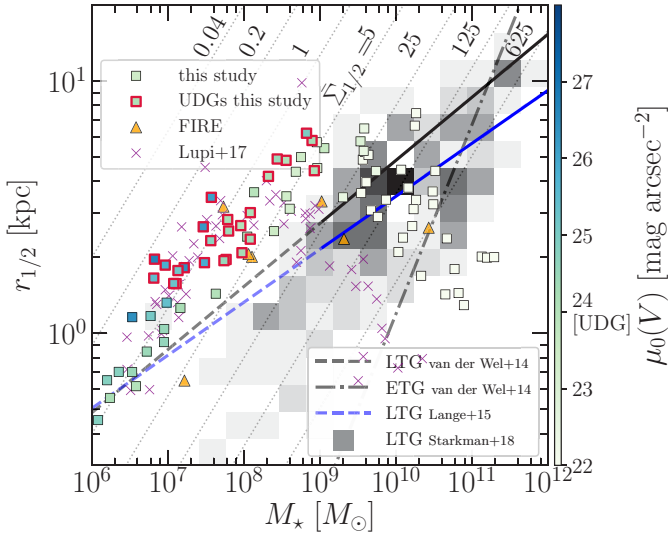} 
\caption{Size versus stellar mass of the NIHAO simulations (squares) at $z=0$.
The facecolours of the squares reflect the central surface-brightness in $V$-band, as indicated by the colour bar.
UDGs are highlighted with red egdes. 
The dotted lines are the contours of the average surface stellar density within $\re$.  
Overplotted are the results from other simulations, as indicated, compiled from \citet{elbadry16}, \citet{chan18}, and \citet{lupi17}.  
The lines are the median $\re-\Mstar$ relations from observations of ``normal" galaxies \citep{lange15,vdw14}, with the shallower and steeper ones for late-type and early-type galaxies, respectively. 
The dashed parts of the late-type relations are extrapolations.
 \quad \quad
In NIHAO, UDGs dominate the mass range $\Mstar=10^{7-9}\Msun$, where the other galaxies that are not UDGs are also quite extended. 
The `bump' in size at $\Mstar\sim10^8\Msun$ is shared by the different simulation suites, which all fail to produce average-sized dwarfs and compact dwarfs.
The greyscale represents the distribution of the SPARC sample of 175 nearby late-type galaxies \citep{starkman18}, highlighting the fact that in the regime of $\Mstar=10^{7-9}\Msun$, there are many relatively compact galaxies observed that are not reproduced by the simulations. 
}
\label{fig:selection} 
\end{figure}

\smallskip
Haloes are identified with the Amiga Halo Finder ({\tt AHF}) \citep{ahf} with virial overdensities 200 times the critical density of the Universe. 
Merging histories are extracted using the complementary {\tt Mergertree} tool of {\tt AHF} from 64 outputs equally spaced in scalefactor between $z\simeq17$ and $z=0$. 
The {\it main progenitor} of the $i$th satellite is defined as the progenitor ($j$) with the maximum figure of merit, $\mathcal{M}=N^2_{i\cap j}/(N_i N_j)$, where $N_i$ and $N_j$ are the number of particles of $i$ and $j$, and $N_{i\cap j}$ is the number of particles shared by $i$ and $j$. 

\smallskip
We compute the magnitudes of star particles in $B$, $V$, and $R$ bands using the Padova simple stellar populations \citep{marigo08ssp} implemented in the software {\tt pynbody} \citep{pynbody}.
We compute the star formation rate by SFR=$\langle \Mstar(t_{\rm age} < t_{\rm max})/t_{\rm max}\rangle_{t_{\rm max}}$, where $\Mstar(t_{\rm age} < t_{\rm max})$ is the mass at birth in stars younger than $t_{\rm max}$, and the average $\langle \cdots \rangle_{t_{\rm max}}$ is obtained by averaging over all $t_{\rm max}$ in the interval $[50, 100]$ Myr in steps of 0.5 Myr. 
The $t_{\rm max}$ in this range are long enough to ensure good statistics. 

\smallskip
Throughout, we consider the effective radius $\re$ as the radius of the sphere that encloses half of the stellar mass, while we have also verified that using the effective radius from fitting a single-S\'ersic component to the $V$-band surface brightness profile does not alter any of our results qualitatively. 
We compute the central surface brightness using stars within the inner $0.25\re$. 
We consider UDGs as galaxies having $\re>1.5\kpc$ and the central surface brightness in $V$-band $\mu_0(V)>24$ mag arcsec$^{-2}$.

\smallskip
We characterize the shape of a system through its shape tensor \citep{allgood06},
\be
\mathcal{S} = \frac{1}{M} \sum_{k} m_k(\mathbf{r}_k)_i (\mathbf{r}_k)_j,
\ee
where $m_k$ is the mass of the $k$th particle, $(\mathbf{r}_k)_i$ is the distance from the centre to the $k$th particle along the axis $i$, and $M$ is the total mass of the volume of interest. 
The eigenvalues of $\mathcal{S}$ are proportional to the squares of the semi-axes ($a>b>c$) of the ellipsoid that describes the spatial distribution of the particles of interest. 
In practice, we consider the the shape of the stellar distribution inside a spherical region of size $\re$, and compute the eigenvalues using an iterative algorithm described in \cite{tomassetti16}. 
We measure the S\'ersic indices $\nS$ by fitting a single S\'ersic component to the stellar surface-density profiles that are obtained by projecting the spherical region within $0.2\Rv$ along a line-of-sight.
The $\nS$ values that we report in the following are measured face-on, i.e., with the projection along the minor axis of the shape tensor. ( We have verified that different projections yield $\nS$ that differ by up to only $\sim$20 percent. )
We define virial radius $\Rv$ as the radius within which the average total density is 200 times the critical density $\rho_{\rm crit}$ of the Universe. 

\section{UDGs in the field}
\label{sec:field}

\begin{figure*}
\includegraphics[width=0.95\textwidth]{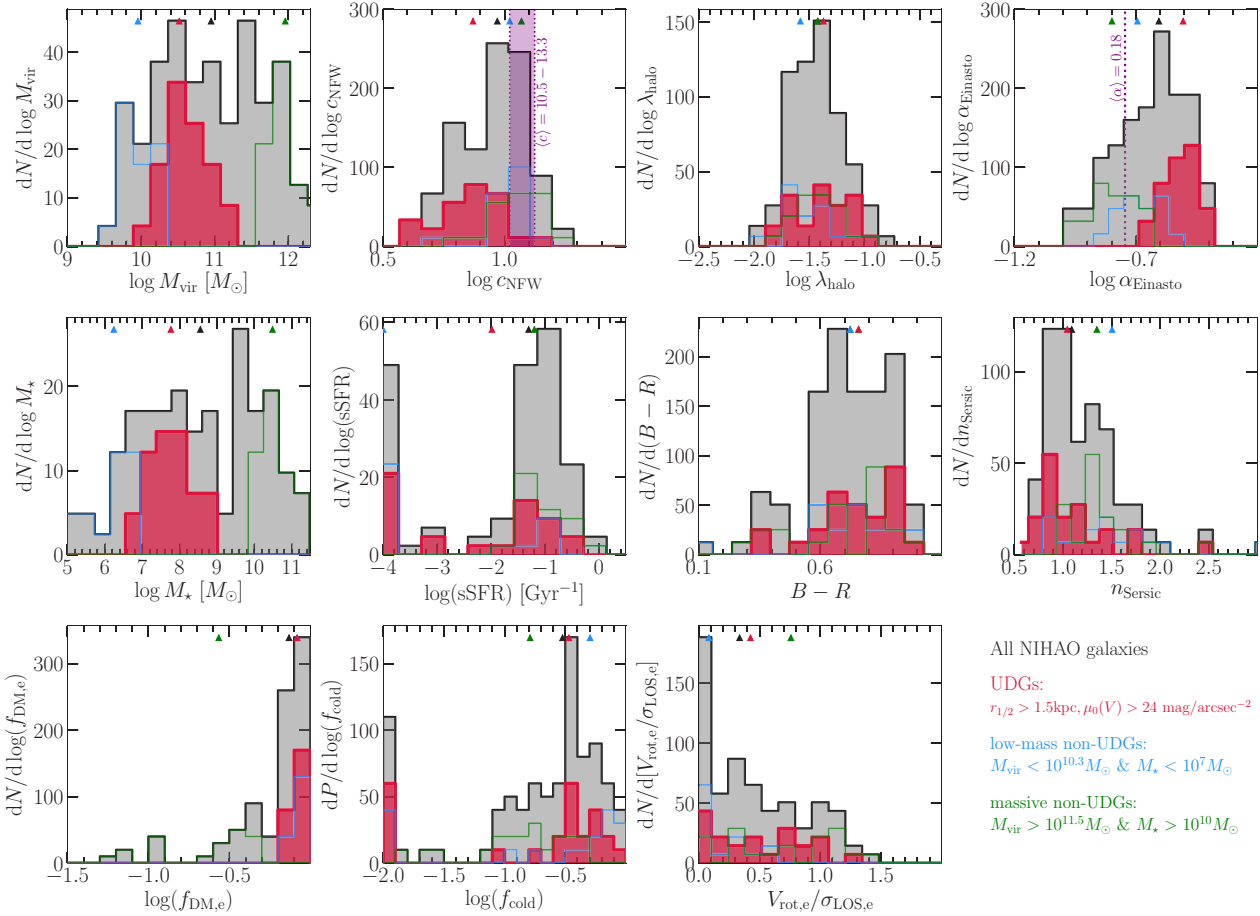} 
\caption{ 
Properties of field UDGs --
The histograms of a handful of properties of the UDGs (red) are contrasted to those of the full NIHAO sample (grey), the massive non-UDGs ($\Mv>10^{11.5}\Msun$ and $\Ms>10^{10}\Msun$; green), and the low-mass non-UDGs ($\Mv<10^{10.3}\Msun$ and $\Ms<10^7\Msun$; blue), with the triangles indicating the medians. 
\quad
{\it Halo properties (top row):} The host haloes of UDGs occupy a narrow mass range ($\Mv\simeq10^{10-11.2}\Msun$); show no obvious difference in the spin parameter ($\lambda_{\rm halo}$); and have lower NFW concentration ($c_{\rm NFW}$) and higher Einasto shape ($\alpha_{\rm Einasto}$) than the non-UDGs and than the median values predicted by $N$-body cosmological simulations for the UDG halo mass range (purple band or line, see text). 
The Einasto shape parameter $\alpha_{\rm Einasto}$ measures the curvature of the logarithmic density profile -- it is $\sim0.18$ for NFW profiles, and is larger for sharper transition from the inner slope to the outer slope.
\quad
{\it Baryonic properties (middle and bottom rows):}
UDGs lie in the stellar mass range of $\Ms\simeq10^{7-9}\Msun$, have S\'ersic indices ($\nS\sim 1$) lower than non-UDGs, but are not fast rotators, with the ratio of rotation and radial velocity dispersion measured at effective radii ($V_{\rm rot,e}/\sigma_{\rm r,e}$) lower than that of massive non-UDGs on average. 
UDGs span a wide range of colour ($B-R$) and specific star formation rate (sSFR), with $\sim30$ per cent being quiescent (which are manually assigned $\log({\rm sSFR}/{\rm Gyr^{-1}})=-4$).
The distribution of the cold gas ($<1.5\times10^4$K) fraction, $f_{\rm cold}\equiv M_{\rm cold}/(M_{\rm cold}+\Ms)$, is similar to that of the sSFR -- the quiescent population is gas-poor (with gas-less systems manually assigned $f_{\rm cold}=0.01$), while the star-forming ones are gas-rich, with $f_{\rm cold}\ga0.4$.
UDGs are the most centrally dark matter-dominated systems, with the dark-matter mass fraction within the effective radius $f_{\rm dm,e}\ga80\%$. 
\quad
Some of the panels in this figure visualise the results reported in Table 1 of \citet{dicintio17}.
The dark-matter halo properties here are measured in the full-physics runs of the NIHAO simulations, while \citeauthor{dicintio17} reported values from the matching dark-matter-only runs. 
}
\label{fig:distributions} 
\end{figure*}
\begin{figure*}
\includegraphics[width=0.65\textwidth]{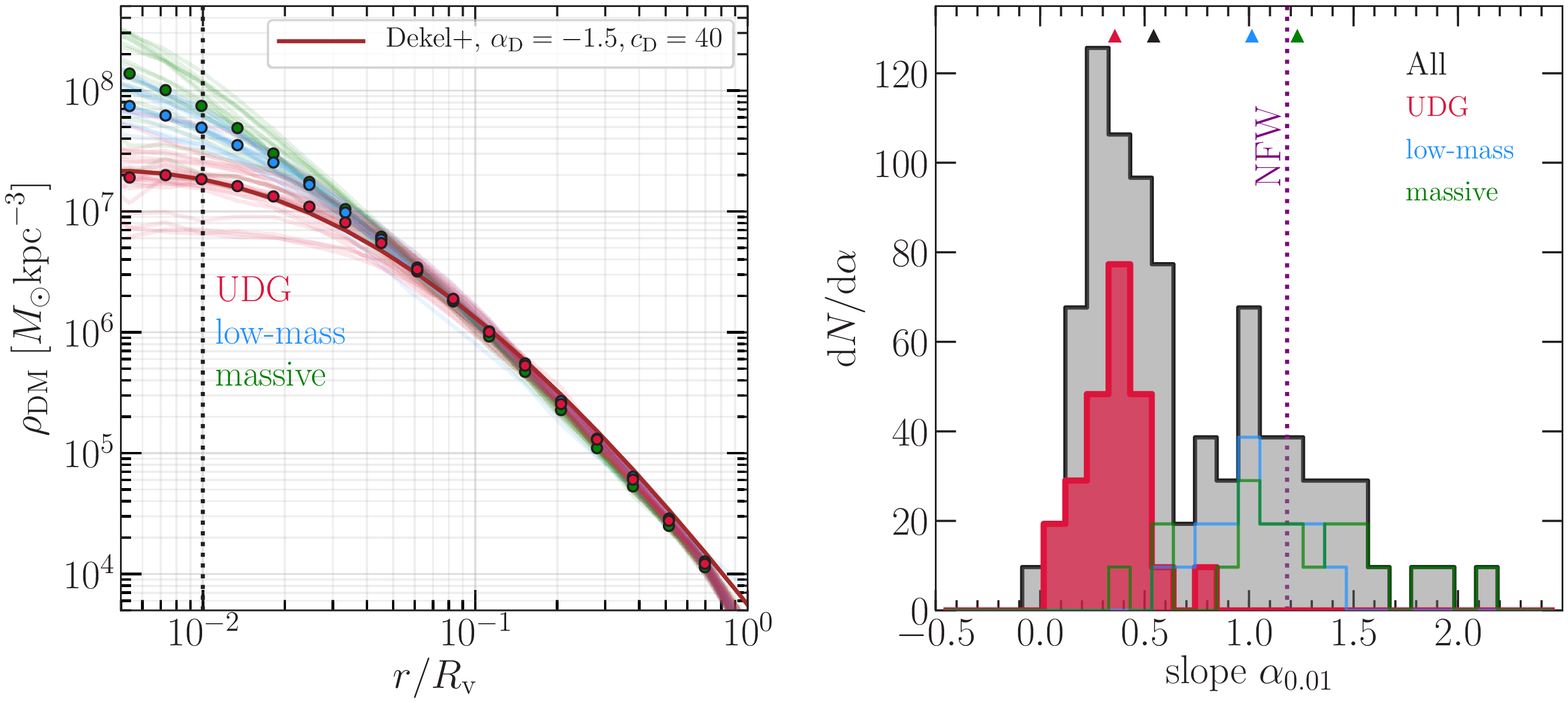} 
\caption{ 
{\it Left:} Density profiles of the host haloes of UDGs.
Thin lines represent individual galaxies; symbols represent the medians; the thick line represents a \citet{dekel17} profile (\equ{dekel}) with $\aD=-1.5$ and $\cD=40$, which approximates the medians the UDGs, and shows a flat dark-matter core. 
\quad
{\it Right:} Histograms of the logarithmic density slope, $\alpha=-\rmd\log\rho/\rmd\log r$, at $r=0.01\Rv$.
The nomenclature and colour scheme are described in \fig{distributions}.
\quad
UDGs show prominent dark-matter cores.
}
\label{fig:core} 
\end{figure*}

\smallskip
\fig{selection} presents the NIHAO sample at $z=0$ on the size-stellar mass plane. 
As can be seen, UDGs lie in a characteristic stellar mass range, $\Mstar=10^{7-9}\Msun$.
In fact, this mass range is dominated by diffuse systems, such that the ``normal'' galaxies in the regime are also quite extended, marginally missing the UDG criteria.
Some of the non-UDGs have similar stellar surface density $\bar{\Sigma}_{\rm 1/2}$ to the UDGs, but are marginally brighter due to their younger stellar populations.
Compared to the observed median size-mass relations \citep{vdw14,lange15}, the simulated galaxies are a factor of $\sim2$ larger in the range $\Mstar=10^{7-9}\Msun$.
However, at $\Mstar<10^7\Msun$ and $>10^9\Msun$, galaxies agree well with the observational relations, yielding a hump of galaxy size right at the UDG scale. 
\cite{dicintio17} suggested that the formation of the UDGs in the NIHAO simulations are associated with bursty star formation histories and therefore episodic SNe outflows.
\cite{dutton16} argued that the response in galaxy size (and host halo structure) to SNe outflows is a function of $\Mstar$, with the regime of maximal  expansion coinciding with the UDG scale.

\smallskip
While the median observational $\re$-$\Ms$ relations that we quote here are based on massive galaxies (with $\Mstar\ga10^9\Msun$) and are extrapolated down to the smaller UDG mass scale, observations have assured the existence of compact dwarf galaxies within the UDG mass range that lie well below the extrapolated $\re$-$\Ms$ relations (e.g., \citealt{norris14}). 
The greyscale in \fig{selection}, for example, represents the distribution of 175 nearby late-type galaxies from the SPARC sample \citep{lelli16}. 
Clearly there is a significant population of dwarf galaxies in the UDG mass range but are quite compact, with $\re\ll1.5$ kpc. 
Therefore, the mass range $\Mstar=10^{7-9}\Msun$ witnesses a variety of structures ranging from the most compact and the most diffuse systems, which is not reproduced by the simulations. 

\smallskip
The problem is actually quite generic in modern simulations with strong stellar feedback.
\fig{selection} also shows a sample of galaxies from the FIRE cosmological simulations, compiled from \cite{chan18} and \cite{elbadry16}, as well as a sample of galaxies from the simulation of \cite{lupi17}.
These simulations differ in various aspects, including subgrid physics and numerical resolution, but all fail in producing compact dwarfs, and exhibit a hump with respect to the extrapolated size-mass relations.
We note that \cite{bose18} showed that in simulations of low gas density threshold for star formation ($n_{\rm th}\sim 1{\rm cm}^{-3}$), the star formation histories of dwarf galaxies are less bursty, the sizes are somewhat smaller than those in NIHAO or FIRE, and there are almost no dark matter cores. 
However, the distributions of the $\re$ of dwarf galaxies in their simulations are still too narrow to capture the observed structural variety.
Therefore, the challenge of $\Lambda$CDM simulations is not in producing UDGs, but in producing compact dwarfs or the diversity of dwarf-galaxy structures. 
With this caveat in mind, we note that UDGs in \fig{selection} are not distinctively special in terms of size or diffuseness compared to the rest of the simulated galaxies of similar stellar mass. 

\subsection{Are (field) UDGs special?}
\label{sec:distributions}

\smallskip
We compare the UDGs with galaxies that are more massive or less massive, in order to see if they are distinctive in certain parameter spaces (other than mass).  
In particular, we define a {\it low-mass} control sample as the non-UDGs with $\Mv<10^{10.3}\Msun$ and $\Ms<10^7\Msun$; and a {\it massive} control sample as those with $\Mv>10^{11.5}\Msun$ and $\Ms>10^{10}\Msun$. 
The massive sample consists of basically $L^\star$ galaxies at $z=0$.

\smallskip
\fig{distributions} contrasts the UDGs with the control samples regarding the distribution function of a collection of properties. 
We can see that the host haloes of UDGs lie in a narrow mass range of $\Mv=10^{10-11.2}\Msun$, clearly lower than the $L^\star$ regime, confirming what is found by \cite{dicintio17}. 
The UDGs do not particularly occupy the high halo-spin tail -- in fact, the spin parameters are distributed similarly to the other galaxies\footnote{We adopt the \cite{bullock01} definition for the spin parameters. The spin parameters, together with all the dark-matter halo properties presented here, are measured in the hydrodynamical NIHAO simulations, while we have confirmed that the result holds qualitatively if we measure them in the companion $N$-body simulations with the same initial conditions.}, with a median of $\langle\lamh\rangle=0.043$.
While the spin-parameter distribution of the UDGs is not special, the \cite{nfw} concentration parameters are on average lower than those of both the low-mass sample and the $L^\star$ galaxies. 
With a median of $\langle\cNFW\rangle =7.3$, the concentration of UDG haloes is significantly lower than what is expected for haloes of the same mass ($\Mv=10^{10-11.2}\Msun$) in $N$-body simulations, which have $\langle\cNFW\rangle\simeq10.5-13.3$ according to the concentration-mass relation of \cite{dm14}. 
Related, the \cite{einasto} shape parameters, $\aE$, of the UDGs are on the higher end, with a median of $\langle \aE\rangle =0.32$.
The shape parameter describes the curvature of the logarithmic density profile, with haloes that obey NFW profiles having $\aE\simeq0.18$.
A higher $\aE$ manifests a sharper transition between the inner and outer logarithmic density slopes than that of a NFW profile (e.g., \citealt{ludlow13}).
The peculiarity of  $\cNFW$ and $\aE$ of the UDGs implies that their host haloes have responded dramatically to baryonic processes, and have dark-matter density profiles significantly different from the NFW form.

\smallskip
\cite{dicintio17} showed that the formation of the field UDGs are associated with bursty star formation histories, which result in episodic, impulsive SNe outflows.
The SNe outflows are believed to be responsible for the cusp-to-core transformation of dark-matter profiles (e.g., \citealt{pg12}).
\fig{core} compares the dark-matter density profiles of the UDGs to those of the low-mass and $L^\star$ samples. 
We see that UDGs exhibit a prominent dark matter core, and find that their density profiles are well approximated by a profile \citep{dekel17} that features flexibility at small $r$, 
\be \label{eq:dekel}
\rho(r) = \frac{\rho_0}{(r/\rs)^\aD [1+(r/\rs)^{1/2}]^{2(3.5-\aD)}},
\ee
where $\rho_0 = (3-\aD) \Mv / 4 \pi \rs^3 g(\cD,\aD)$, $g(x,\aD)\equiv x^{3-\aD} / (1+x^{1/2})^{2(3-\aD)}$, and $\rs=\Rv/\cD$.
The profile is defined by three parameters: the virial mass, $\Mv$, the logarithmic density slope at $r\to0$, $\aD$, and a concentration parameter, $\cD$.
On average, the UDGs are well described by \equ{dekel} with $\aD=-1.5$ and $\cD=40$.
The right-hand panel of \fig{core} shows the distribution of the logarithmic density slope $\alpha = -\rmd\log\rho/\rmd\log r$ evaluated at $r=0.01\Rv$.
Most of the UDGs lie in the narrow range of $\alpha_{0.01}=0-0.5$.
\cite{dicintio14a,dicintio14b} and \cite{tollet16} expressed the response in halo profile to SNe feedback as a function of the star-formation efficiency $\Ms/\Mv$, and we have verified that the cores in the UDG hosts are consistent with their empirical relation.
In a companion study (Freundlich et al. in prep), we link the gas inflows and outflows in the central regions of the NIHAO galaxies to the changes of halo density profiles using a simple analytic model that makes use of \equ{dekel}. 

\smallskip
Despite having dark-matter cores, the UDGs are among the most dark matter dominated systems: their dark-matter mass fractions within the effective radius ($f_{\rm dm,e}$) are typically over $80$ per cent. 

\smallskip
Regarding the baryonic properties, the UDGs have a median S\'ersic index of $\simeq1$, showing a mode of the $\nS$ distribution at $\simeq0.8$, lower than that of the non-UDGs.
The low S\'ersic indices do not mean that UDGs are flattened, rotation-supported systems. 
In fact, the UDGs are not fast rotators, with the ratios of rotation speed to the radial velocity dispersion ($v/\sigma$) at $\re$ similar to those of the full sample.
As we will see shortly, the UDGs are mostly {\it not} oblate in shape.

\smallskip
The UDGs show a wide range of sSFR and colour.
While the $L^\star$ analogues are mostly star-forming, with sSFR$>0.01$Gyr$^{-1}$, the UDGs seem to show a bimodality in sSFR, and are overall slightly redder.  
About 30 per cent of the field UDGs are not forming stars instantaneously at $z=0$. 
The star-forming UDGs have modestly high cold gas fractions, with $f_{\rm cold}\equiv M_{\rm cold}/(\Ms + M_{\rm cold})\ga 0.4$, which is consistent with those of a few observed field UDGs \citep{papastergis17}.

\subsection{Morphology and shape}
\label{sec:shape}

\begin{figure}
\includegraphics[width=0.5\textwidth]{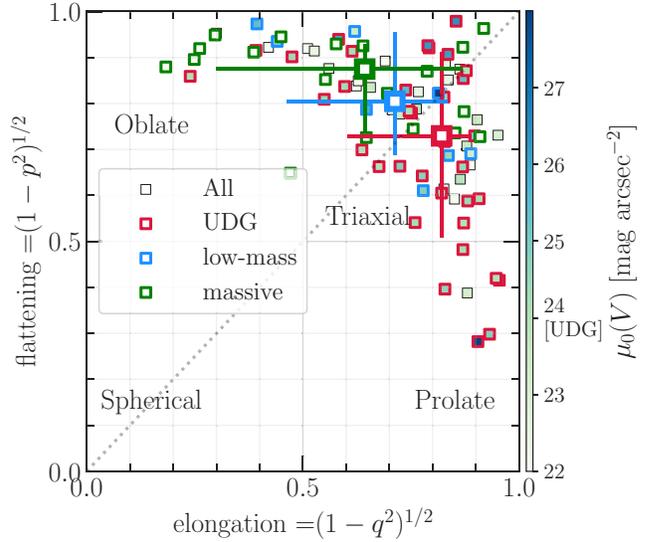} 
\caption{ 
Shape of field UDGs -- `flattening' $(1-p^2)^{1/2}$ versus `elongation' $(1-q^2)^{1/2}$, where $p=c/b$ and $q=b/a$ are the axes-ratios, and $a$(set to be $\re$), $b$, and $c$ are the lengths of the semi-axes of the eigen-ellipsoid that describes the stellar distribution within $\re$, with $a>b>c$.
In this space, the four categories of shapes are separated into the four quadrants, as indicated.
}
\label{fig:shape} 
\end{figure}

\smallskip
We characterize the 3D shape of a galaxy by introducing the elongation ($e$) and flattening ($f$) parameters, defined as $e=(1-b^2/a^2)^{1/2}$ and $f=(1-c^2/b^2)^{1/2}$, respectively, where $a$, $b$, and $c$ ($a>b>c$) are the lengths of the semi-axes of the eigen-ellipsoid describing the stellar distribution within the half stellar mass radius.
\footnote{ We have set the major axis $a$ to be the half stellar-mass radius $\re$.}
In the space spanned by $e$ and $f$, galaxies of different shapes, namely, oblate ($a \ga b\gg c$), triaxial ($a>b>c$), prolate($a\gg b\ga c$), and spherical ($a \ga b \ga c$) systems, are well separated into four quarters, as shown in the left-hand panel of \fig{shape}.
While the majority of the NIHAO galaxies are triaxial, the UDGs are significantly more prolate than the $L^\star$ galaxies.
The UDGs are also slightly more prolate than the low-mass dwarfs.
In fact, in the prolate quarter of the parameter space, most of the systems are UDGs.

Our result is qualitatively consistent with the observational result of \cite{burkert17}, who tested two simple scenarios about the intrinsic shapes of UDGs by comparing their predictions of the apparent axis-ratio distribution with what is observed in the Coma cluster \citep{koda15}.
\cite{burkert17} assumed that UDGs are either perfectly oblate ($a=b>c$) or perfectly prolate ($a>b=c$), with different axis-ratios, $q=c/a$, and are observed at uniformly random viewing angles.
They showed that the observed apparent axis-ratio distribution is compatible with the all-prolate scenario, while the all-oblate scenario over-predicts the abundance of systems that appear round. 

\smallskip
We also note that, in the context of high-redshift galaxies, the galaxies of masses below $\Ms \simeq10^{9.5}\Msun$ tend to be prolate when they are dark-matter dominated in the centre \citep{tomassetti16, ceverino15}.
While most of the UDGs are at discovered at $z\simeq0$, their triaxiality/prolateness and  high dark-matter fraction $f_{\rm dm,e}$ fit consistently in this picture.




\section{Satellite UDGs}
\label{sec:group}

\begin{figure}
\includegraphics[width=0.45\textwidth]{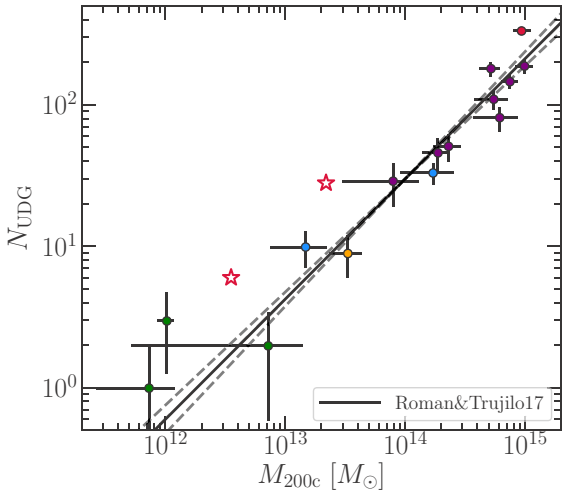} 
\caption{ 
Abundance of satellite UDGs down to $\mu_0(V)\approx29$mag arcsec$^{-2}$, as a function of the virial mass of the host group/cluster.
The upper red star represents the galaxy group used in this study; the lower red star represents the most massive galaxy in the NIHAO suite, which is of the mass scale of a compact-group.
The circles with errorbars are observations compiled by \citet{rt17} (see references therein).
The lines represent $N_{\rm UDG}\simeq 30 (\Mv/10^{14} \Msun)^{0.85\pm0.05}.$ 
\quad 
The numbers of UDGs in the two simulated groups seem to be on the high side but in the ballpark of the observations.   
}
\label{fig:sanity} 
\end{figure}

\smallskip
In this section, we focus on the UDGs in a simulated galaxy group with $\Mv($z=0$)=10^{13.33}\Msun$ (``halo 4.2" in \citealt{dutton15}). 
The host halo has a virial radius of $\Rv=572$ kpc, and the bright central galaxy has a stellar mass of $\Ms=10^{11.4}\Msun$ at $z=0$.
As a sanity check of whether the simulations produce a reasonable amount of satellite UDGs, we show in \fig{sanity} the number of UDGs versus the host halo mass, comparing the simulations with the observations compiled by \cite{rt17}. 
The data are complete down to a surface brightness of $\mu_0(V)\simeq29$mag arcsec$^{-2}$, so we have applied the same surface brightness cut for the simulations.  
There are two simulation results here: in addition to the ``halo4.2'' that we will analyse, the other is the most massive system from the NIHAO suite, with a virial mass of $\Mv\approx10^{12.5}\Msun$. 
Previously, we have used the central galaxy of this system, while here we count its satellite UDGs.
The numbers of UDGs in the simulations are on the high side, but lie in the ballpark of the observational estimates, given the large errors in $\Mv$.   
We caution again that, the simulations easily produce UDGs but hardly any compact satellite. 

\subsection{Radial trends}
\label{sec:radial}

\begin{figure*}
\includegraphics[width=0.67\textwidth]{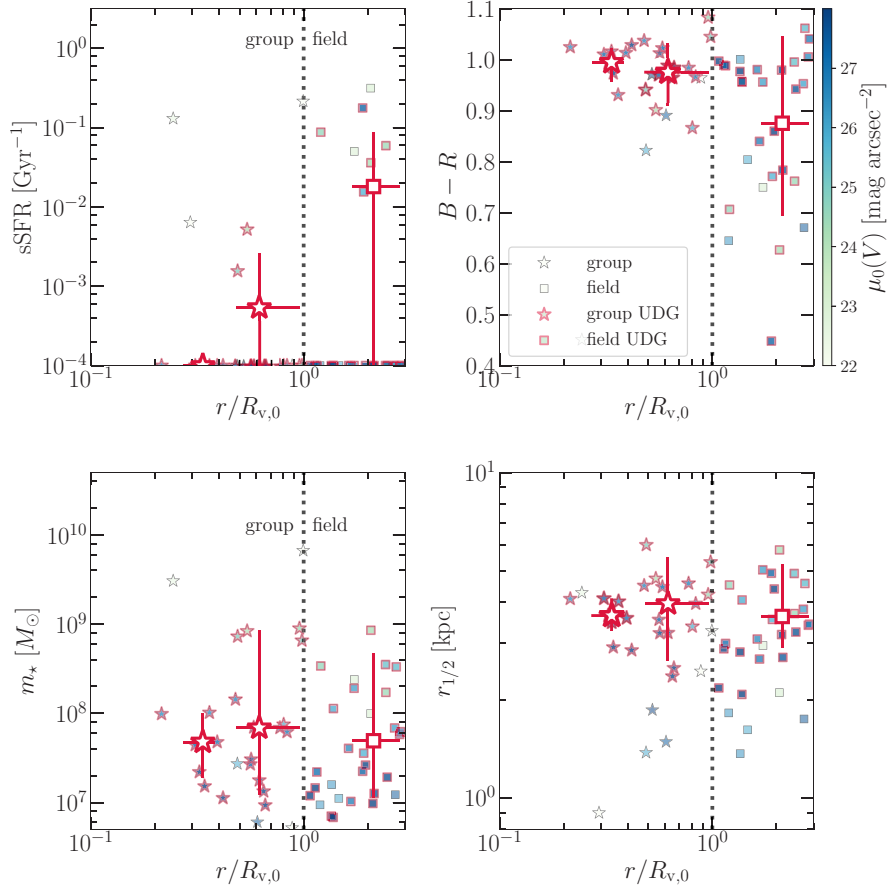} 
\caption{ 
UDG properties versus group-centric distance $r$ (in units of the present-day $\Rv$ of the galaxy group).
Each symbol represents a galaxy in the galaxy-group simulation at $z=0$. 
The UDGs are highlighted with red edges.
Stars and squares represent galaxies inside and outside $\Rv$, respectively.  
The large symbols indicate the medians of the UDGs in three radial bins: $r/\Rv\in(0.3,0.5), (0.5,1)$, and $(1.5,3)$, and the errorbars represent the 16th and 84th percentiles. 
\quad
UDGs exhibit a wide range of sSFR and colours outside $\Rv$, and are quiescent and red in the group, as observed. 
Their stellar mass $\mstar$ and the effective radius $\re$ do not show obvious radial trends.
}
\label{fig:QversusRadius} 
\end{figure*}

\smallskip
Since UDGs are observed both in the field and in clusters and groups, an intuitive scenario for the formation of satellite UDGs is that they were already puffed up when in the field and became quenched after falling into a dense environment, as implied by several studies (e.g., \citealt{rt17}, \citealt{alabi18}, \citealt{fm18}, and \citealt{chan18}).
Recent observations seem to support the aforementioned scenario -- 
\cite{rt17}, using 11 UDGs near a few compact galaxy groups, showed that the red ones are predominantly located at projected distances less than 200 kpc (i.e., $\la \Rv$) from the group centres; 
\cite{alabi18}, using 16 UDGs with spectroscopically confirmed Coma-membership, found that those at smaller projected cluster-centric distances are redder . 

\smallskip

\smallskip
\fig{QversusRadius} presents the properties of the group UDGs as a function of the 3D host-centric distance.
The radial trends are in good qualitative agreement with the observations --  in the inner part of the galaxy group, the UDGs are almost exclusively quiescent and red; towards the outskirts, star-forming and bluer UDGs start to exist; outside $\Rv$, the UDGs exhibit a wide range of colours, consistent with the results for field UDGs shown in \fig{distributions}.

\begin{figure*}
\includegraphics[width=0.7\textwidth]{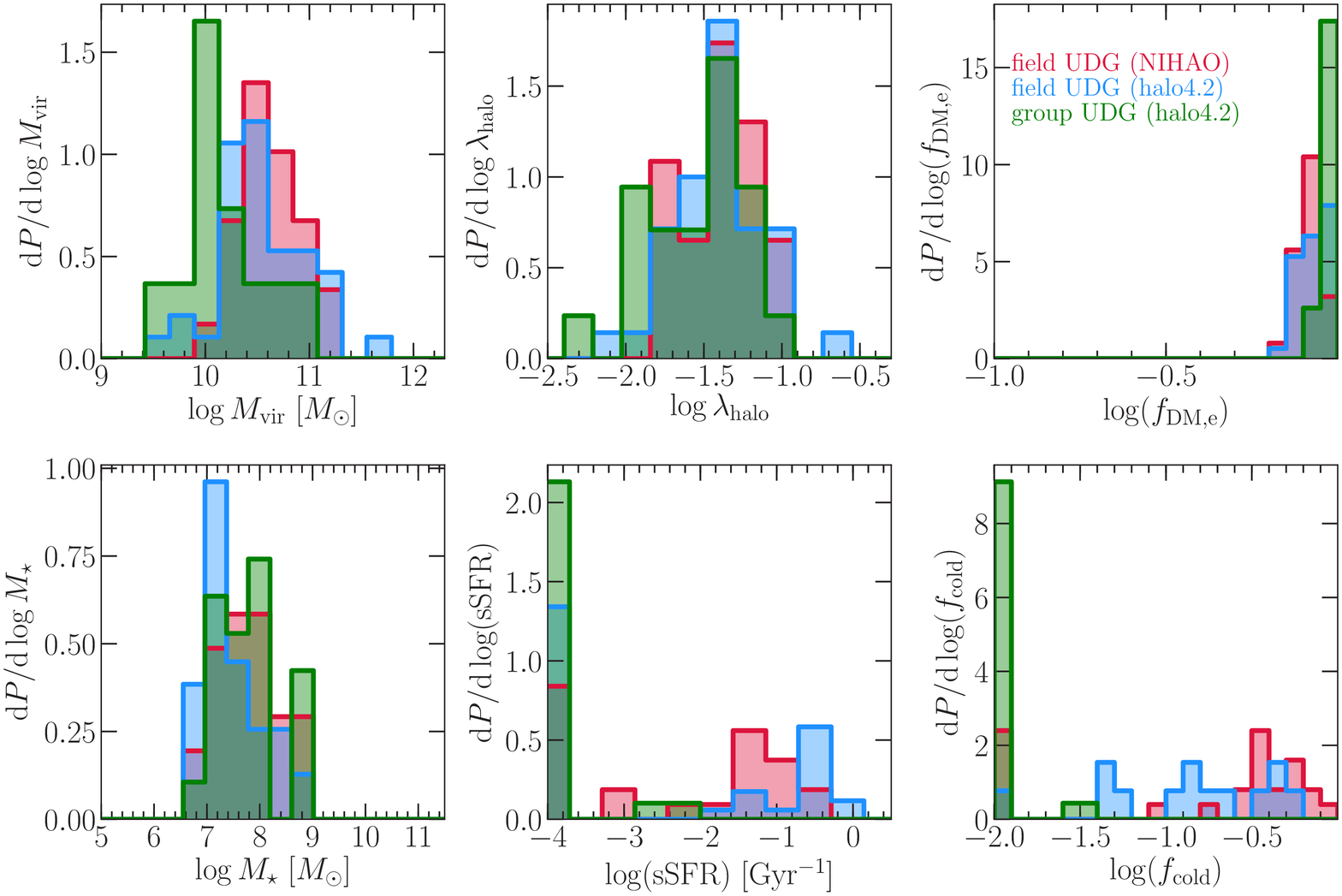} 
\caption{ 
Comparison of the properties of field UDGs in the NIHAO suite and in the simulation of the group-sized system ``halo4.2''.
The field UDGs in the NIHAO simulations (red) and in ``halo4.2'' (blue) have similar properties regarding halo mass ($\Mv$), stellar mass ($\Ms$), halo spin ($\lamh$), specific star formation rate (sSFR), dark-matter fraction within half stellar-mass radius ($f_{\rm DM,e}$), and cold-gas ($<1.5\times10^4$K) fraction ($f_{\rm cold}$). 
The group UDGs in ``halo4.2'' (green) are mostly quenched and cold-gas poor, and have marginally lower halo mass than their counterparts in the field.
}
\label{fig:QdistributionComparison} 
\end{figure*}

\smallskip
\fig{QdistributionComparison} presents a comparison of UDGs in the NIHAO simulations and in the simulation of the group-sized system `halo4.2'.
The aims of the comparison are twofold.
First, there are two samples of field UDGs in this study, the UDGs outside virial radius of the group, and the UDGs in the NIHAO simulations -- a comparison of them serves as a check of what we have learnt about field UDGs using the NIHAO simulations.
Reassuringly, despite the fact that the group simulation has poorer numerical resolution than the NIHAO simulations, the two field-UDG populations share similar global properties.
Their halo mass and stellar mass lie in the range of $10^{10.5\pm0.6}$ and $10^{7.8\pm1.0}$, respectively; their host-haloes show ordinary spin distribution; they are both dark-matter dominated in the centre; they both show a wide range of specific star formation rate, with about 30 per cent quiescent; and they both show a significant fraction of cold gas. 
(It is beyond the scope of this paper to determine the origins of the quiescence in the UDGs at large group-centric radii. 
We speculate though some of the non-star-forming UDGs may be splashback satellites, associated with other groups, or temporarily exhausted in gas.)
Second, we compare the group and field populations, and confirm what is shown in the upper panels of \fig{QversusRadius}, that the group UDGs are mostly quiescent and gas-poor. 
There is a weak trend that the group UDGs have lower halo mass, manifesting tidal stripping of dark matter mass. 

\smallskip
In this scenario, what causes UDGs to lose their gas reservoir in the host system is either tidal stripping or ram pressure stripping. 
The question is which quenching mechanism is dominant. 
If tidal stripping is more important, the tidal force that strips the cold gas can also remove the stars from the outskirts of the satellite, reducing its $\re$ and $\Ms$. 
Since the tidal field is stronger in the inner part of a group, an inevitable side effect, if tidal stripping is the dominant quenching mechanism, is that the UDGs closer to the cluster centre would have lower $\Ms$ and $\re$ than the UDGs on the outskirts or in the field. 

\smallskip
Interestingly, as shown in the bottom panels of \fig{QversusRadius}, there is no obvious radial gradient in stellar mass or size, which seems to suggest that tidal stripping is not the dominant factor in quenching the UDGs.
In what follows, we try to rationalize these radial trends, or the lack thereof, by inspecting the evolution of the satellite galaxies. 
We will show that tidal puffing up is partly responsible for the lack of radial trend in $\re$.
From now on, in the context of satellite galaxy evolution, we denote the satellite-centric radius by $l$, and host-centric distance by $r$.

\subsection{Evolution}
\label{sec:evolution}

\begin{figure*}
\includegraphics[width=0.75\textwidth]{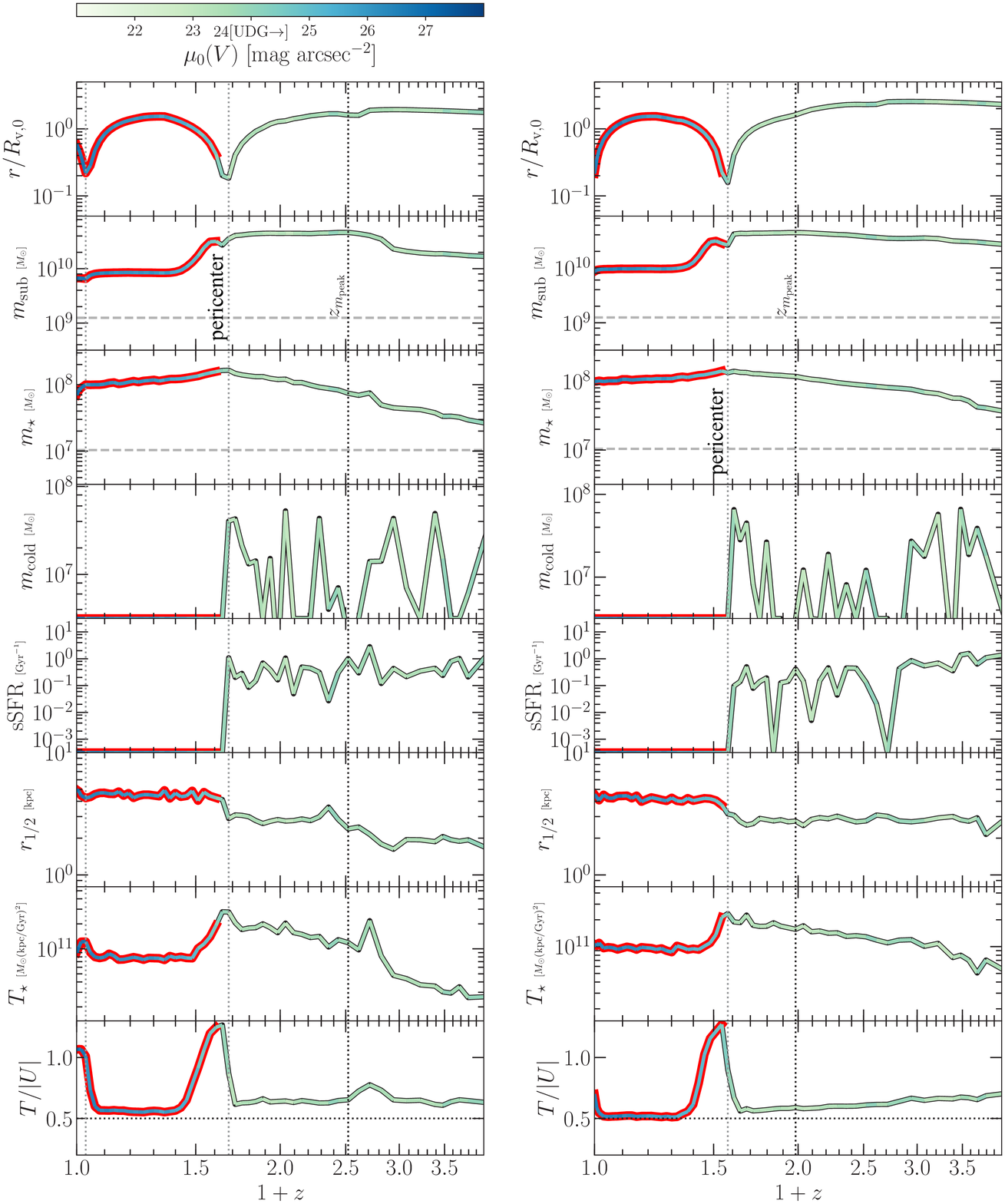} 
\caption{ Evolution of satellite galaxies that become UDGs at orbital pericentre.
Two examples are presented here (on the left-hand side and the right-hand side, respectively), showing the following quantities as functions of redshift -- 
from the top to the bottom --
group-centric distance $r$ in units of the present-day virial radius of the host;
subhalo mass (with the dashed line marking the resolution mass of $\mres = 10^{9.1}\Msun$, which corresponds to 250 dark-matter particles); 
stellar mass (with the dashed line marking the resolution mass of $\mres = 10^{7}\Msun$, which corresponds to $\sim100$ star particles);
the mass of cold gas ($<1.5\times10^4K$);
the specific star-formation rate (sSFR);
the half stellar-mass radius $\re$; 
the kinetic energy in stars $T_\star$;
the ratio of the total kinetic energy to the binding energy $T/|U|$ (with the horizontal dotted line marking the virial-equilibrium value of 0.5).
The facecolour of the lines reflects the $V$-band central surface brightness, as indicated by the colour bar on the top.  
The UDG-phases are highlighted with red edges. 
The thicker vertical line marks the infall redshift, $\zpeak$, when $\msub$ reaches the maximum.
The thin vertical lines indicate the orbital pericentres.
\quad
The satellites are puffed up at the first pericentre passage, becoming UDGs. 
This is accompanied by significant stripping of dark matter mass, negligible change in stellar mass, complete removal of cold gas, a spike in the stellar kinetic energy, and a short deviation from virial equilibrium, all happening within a period of a couple of dynamical times. 
}
\label{fig:Qhistories1} 
\end{figure*}

\smallskip
We begin by showing case-studies of representative satellite galaxies, and then examine the average evolutionary tracks for a statistical sample. 
We define the infall redshift ($\zpeak$) of a satellite galaxy as the redshift when its host subhalo mass reaches the maximum throughout history. 

\subsubsection{Case study: UDGs transformed from ``normal'' dwarfs at pericentre} 

\smallskip
We find a population of satellite galaxies that were not UDGs 
at infall but become UDGs inside the group.
This amounts to 50 per cent of the surviving satellite-UDG population.
\fig{Qhistories1} presents two examples, showing the evolution of a collection of quantities.
The two satellites are both puffed up and become UDGs right after the first pericentre passage, becoming UDGs. 

\smallskip
The expansion at the pericentre is accompanied by a few other changes, including significant dark matter mass loss and a complete removal of cold gas.
The change in stellar mass is small, implying that tidal stripping is marginal inside the baryonic range of the galaxy where stars and cold gas reside.
Given that the cold gas is completely lost at the pericentre, ram pressure seems to be the main cause of the quenching of their star formation.
We will discuss further the roles of tidal stripping and ram pressure stripping in \se{stripping}.

\smallskip
The increase in size also coincides with a spike in the kinetic energy of stars, and a deviation from virial equilibrium of the whole system, as can be seen from the ratio of kinetic energy and potential energy, $T/|U|$. 
These phenomena together are indicative of {\it impulsive tidal heating} -- a process describing what happens when the duration of the encounter of the system of interest (i.e., the satellite) and the perturber (i.e., the centre of the host system) is shorter than the crossing time of the constituent particles within the system of interest. 
During an impulsive encounter, the particles will be given a kinetic energy $\Delta T$ while retaining their potential energy instantaneously; after the satellite relaxes to a new equilibrium state (i.e., when $T/|U|$ drops back to $\sim0.5$), the kinetic energy of the particles will decrease by the amount of $2\Delta T$ (if they are not stripped away); and finally, conserving the total energy, the potential energy of the affected particles increases, resulting in a size growth. 

\smallskip
This picture is manifested exemplarily in \fig{Qhistories1} -- over the period of time between the initial and the new equilibrium states, the kinetic energy of the stars first rises, and then drops to a value that is lower than that before the pericentre encounter, accompanied by the increase in $\re$. 
Therefore, new UDGs can be created out of normal dwarf satellites through tidal heating in a dense environment.
\cite{carleton18} modelled the size evolution of satellite galaxies due to tidal heating using an empirical recipe for tidal evolution from \cite{errani18}, and showed that satellite galaxies in cored DM subhaloes of $\msub=10^{10-11}\Msun$ can become UDGs. 
This is consistent with our finding with the simulations. 
In \se{heating}, we provide further justification of this mechanism, and explore the condition for optimal puffing up. 

\begin{figure*}
\includegraphics[width=0.75\textwidth]{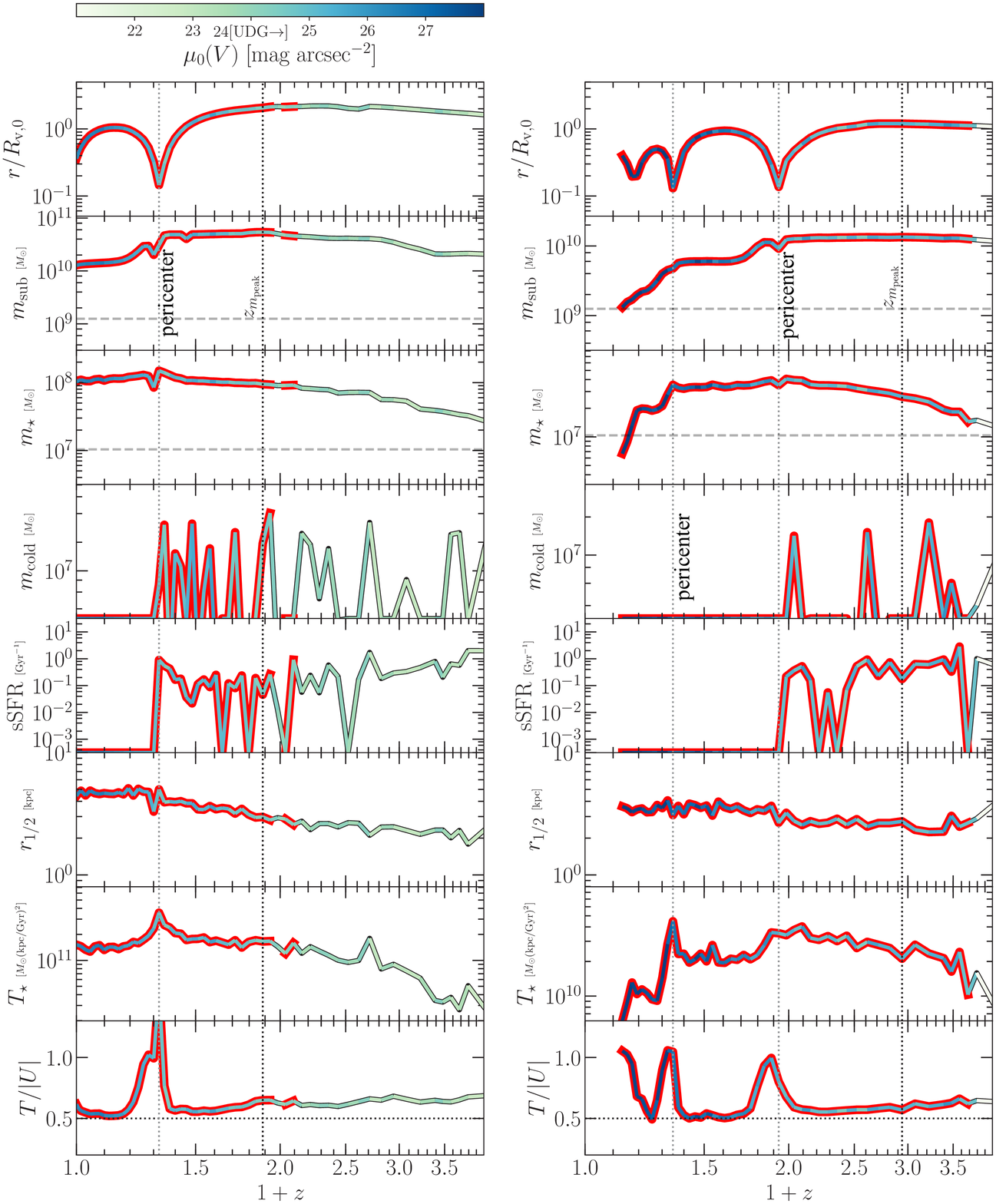} 
\caption{Similar to \fig{Qhistories1}, but showing examples of the evolution of two galaxies that were UDGs already at infall. 
One of them survives to $z=0$ ({\it left-hand column}); the other has been disrupted ({\it right-hand column}).
\quad
Field UDGs can either survive a dense environment or get disrupted. 
}
\label{fig:Qhistories2} 
\end{figure*}
%

\subsubsection{Case study: satellites accreted as UDGs} 

\smallskip
There are also satellite galaxies that were already UDGs at infall. 
Some of them survive the group environment and continue to exist at $z=0$; others have been disrupted or merged into the central galaxy. 
Representative examples are shown in \fig{Qhistories2}. 
We can see that, as long as a UDG survives, it exhibits similar behaviors to the satellites that become UDGs at pericentres discussed previously.
That is, at pericentres, there is a significant $\msub$ decrease and a marginal $\mstar$ change, a complete loss of cold gas, and a size growth together with energetics that are indicative of tidal heating. 
In the case where the UDG is disrupted, as shown in the right-hand panel of  \fig{Qhistories2}, the disruption is preceded by a significant drop in stellar mass, implying that the instantaneous tidal radius is well inside the stellar mass distribution.

\smallskip
These galaxies were already UDGs in the field. 
We {\it assume} that they form in the same way as how the field UDGs in the NIHAO simulations form, i.e., via repeated SNe outflows associated with bursty star-formation histories, although the resolution of the group simulation is not high enough to resolve the dark-matter core formation in detail.
This assumption is supported by the similarity in the global properties of the two field-UDG populations shown in \fig{QdistributionComparison}, and also by their bursty star formation histories shown in \fig{Qhistories2}.
\begin{figure*}
\includegraphics[width=0.75\textwidth]{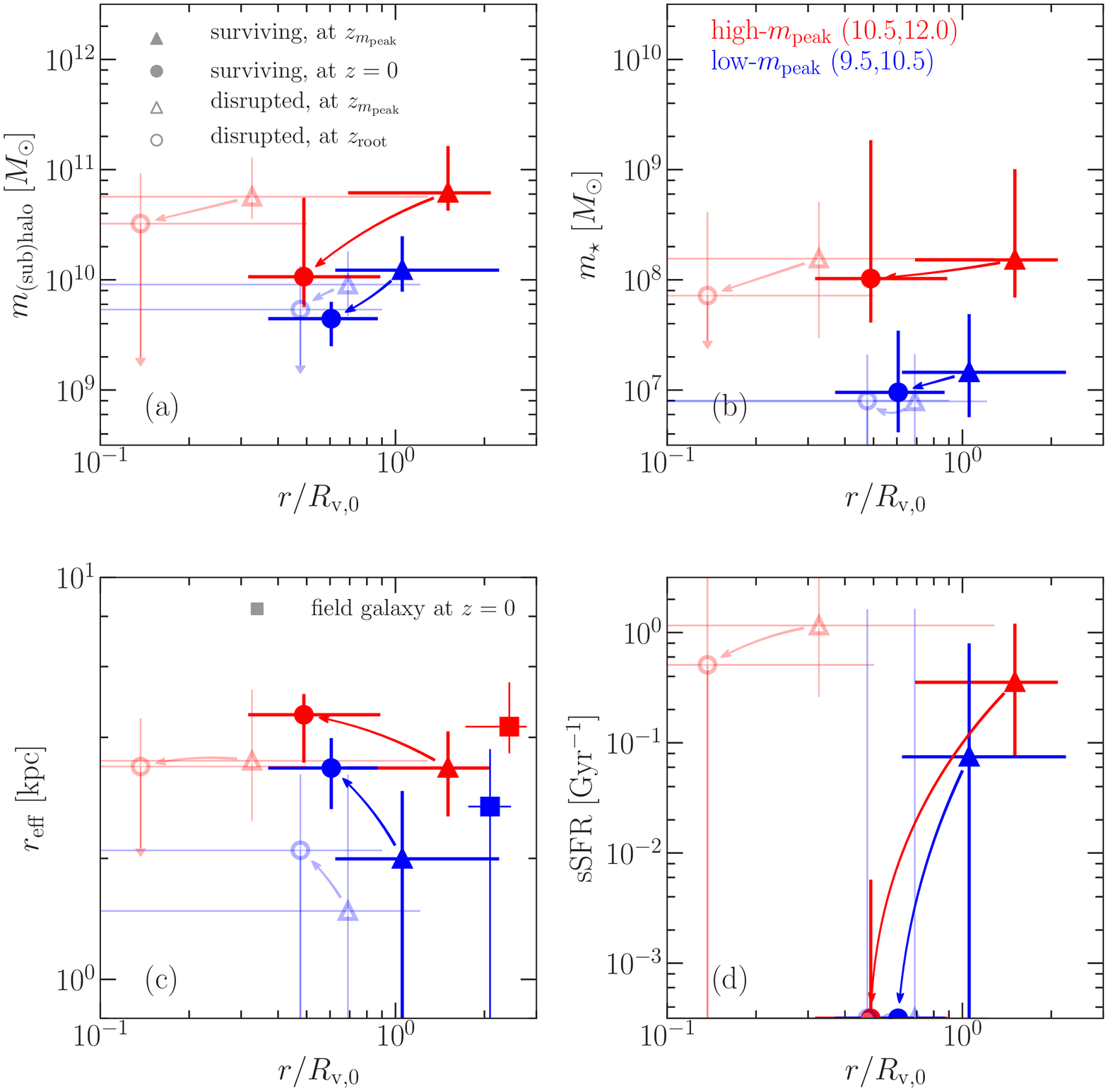} 
\caption{ Satellite properties versus group-centric distance $r$ in units of the present-day virial radius of the host -- showing the average evolution between infall ($\zpeak$) and the lowest redshift when the galaxies are still gravitationally bound and identifiable ($\zroot$). 
The satellites are binned by whether they survive at the present day ($\zroot=0$) or have been disrupted ($\zroot>0$), and by their subhalo mass at infall ($\mpeak$), as indicated. 
The triangles and circles mark the medians at $\zpeak$ and $\zroot$, respectively, connected by arrows indicating the direction of evolution (just for guiding your eyes, not indicating the median values along the evolutionary track).
The bars represent the 16th and 84th percentiles. 
\quad
The $r$-coordinate of a triangle roughly corresponds to the virial radius of the group at the infall time of the satellites -- the open triangles are at smaller $r$ than the solid ones, meaning that the disrupted satellites are accreted earlier than the surviving ones.
The open circles indicate the median locations of the disrupted satellites right before disruption, and are at smaller $r$ than the surviving satellites at $z=0$ (solid circles), meaning that disruption tends to happen closer to group centre.
High-$\mpeak$ systems lose more dark-matter mass and travel to smaller radii than low-$\mpeak$ ones [Panel (a)], reflecting the influence of dynamical friction.
While the stripping of dark-matter mass is significant, the stellar-mass loss is marginal [Panel (b)].
Despite the weak decrease in $\mstar$, satellites generally grow in size in the group [Panel (c)], indicative of tidal heating.
Field galaxies in the vicinity of the group today ({\it squares}) are larger than the satellites of similar $\mpeak$ at accretion ({\it triangles}), but are comparable in size to the puffed up systems at $z=0$ ({\it circles}).
Galaxies become quenched from infall to $z=0$ [Panel (d)]. 
}
\label{fig:QversusRadiusHistories} 
\end{figure*}
%

\subsubsection{Average evolution} 

\smallskip
We now consider the evolution of satellite galaxies statistically, in terms of the quantity of interest ($Q$) versus the distance to the group centre ($r$). 
We split our sample of satellite galaxies by their subhalo mass at infall ($\mpeak$) into two bins -- for each $\mpeak$ bin, we compare, in the $Q-r$ plane, their average locations at infall ($\zpeak$) and at the latest time ($\zroot$) when they are still gravitationally bound and detectable by the halo finder\footnote{For the surviving satellites, $\zroot=0$. For the satellites that have been disrupted or merged with the central, $\zroot$ is the lastest redshift when they still can be detected by the {\tt AHF} halo finder.}.
The results are presented in \fig{QversusRadiusHistories}, for $Q$ being the subhalo mass ($\msub$), the stellar mass ($\mstar$), the effective radius ($\re$), and the specific star-formation-rate (sSFR). 
Several illuminating behaviors are revealed.

\smallskip 
First, we gain some insights into the conditions for disruption.  
Since the virial radius of the host halo grows in time, and since $\zpeak$ is approximately when the first virial-crossing occurs\footnote{Satellite-galaxy progenitors start to lose mass at out to $\sim2\Rv$ from the group centre, consistent with the finding of \cite{behroozi14}.}
, smaller $r$ at $\zpeak$ corresponds to earlier infall. 
Therefore, comparing the horizontal coordinates of the open and solid triangles in (any panel of) \fig{QversusRadiusHistories}, we learn that the disrupted satellites are accreted earlier than the surviving satellites. 
This is not surprising: the systems that have spent longer time in the tidal field of the group and have interacted with the denser core of the host halo when the host was smaller should be more likely to be disrupted. 
Comparing open and solid circles, we can see that disruption generally occurs at smaller group-centric distances than where the surviving satellites are today.
This indicates that systems are more prone to disruption when they are closer to the group centre, where the density is higher and tidal interactions are more intense. 

\smallskip 
Second, we learn from the behaviors of the surviving satellites. 
Panel (a) of \fig{QversusRadiusHistories} shows that the high-$\mpeak$ satellites lose more halo mass than the low-$\mpeak$ ones and end up at marginally smaller group-centric distances. 
Dynamical friction (DF) brings satellites closer to the centre of the host and facilitates tidal stripping, and the timescale of DF is a strong function of the mass ratio of the satellite and the host -- it is longer than a Hubble time for $\msub/\Mv\la0.01$, but decreases sharply with increasing mass ratio (see e.g.,\citealt{taffoni03}, \citealt{bk08}).
Given the virial mass of the host group, $\Mv=10^{13.33}\Msun$, our high-$\mpeak$ and low-$\mpeak$ bins correspond to the two regimes where dynamical friction is (marginally) efficient and inefficient, respectively. 
Therefore, the phenomenon that the high-$\mpeak$ satellites lose more mass and end up closer to the group centre agrees with what is expected from DF (but see also \cite{vdb16}, which showed that more massive satellites have lower specific orbital energy at infall). 

\smallskip 
Panel (b) of \fig{QversusRadiusHistories} shows that the average stellar mass loss between infall and $z=0$ is weak. 
This explains the null radial trend of $\mstar$ with distance $r$ as shown in \fig{QversusRadius}.

\smallskip 
Panel (c) of \fig{QversusRadiusHistories} shows that galaxies generally grow in size in the group environment.
This tells us that the puffing up at orbital pericentres dominates the size evolution over the stripping of stellar mass. 
However, how do we comprehend the null trend of $\re$ with distance $r$ shown in \fig{QversusRadius}?
We think that {\it progenitor bias} helps to complete the story, which states that, for a fixed stellar mass, galaxies are more compact at earlier times due to the universe being denser at higher redshifts. 
As shown by \cite{vdw14}, at fixed $\Ms$, the average size of star forming galaxies scales with redshift roughly as $\re\propto(1+z)^{-0.75}$. 
Taking our high-$\mpeak$ population as an example -- the average infall redshift is $\sim1$, so their sizes at infall is smaller than the size of field-galaxies of similar mass at $z=0$ by about $40$ per cent.
To illustrate this point, we overplot in Panel (c) the average size of the galaxies at $z=0$ that lie in the range of $\Rv <r<3\Rv$ ({\it squares}). 
Indeed, they are larger than the satellites in the corresponding $\mpeak$ bin at infall ({\it triangles}), but similar in size to the evolved satellites at $z=0$ ({\it circles}).

\smallskip 
Finally, Panel (d) shows that the satellites were generally star forming at infall, and are quenched (sSFR$<10^{-2}$Gyr$^{-1}$) at $z=0$. 
Therefore, the radial trend of sSFR and colour of UDGs shown in \fig{QversusRadius} simply reflects environmental quenching of satellites, as anticipated.  

\section{Toy model: the origin of group UDGs}
\label{sec:discussion}

\smallskip 
In the previous section, we used hints from the simulations to suggest that the quenching of group UDGs is due to ram pressure stripping rather than tidal stripping. 
We also raised the possibility that the size increase at orbital pericentres reflects tidal heating.
In this section, we use simple analytic estimates of these two effects in order to evaluate the validity of our conclusions.

\subsection{Tidal stripping versus ram-pressure stripping?}
\label{sec:stripping}

\begin{figure}
\includegraphics[width=0.5\textwidth]{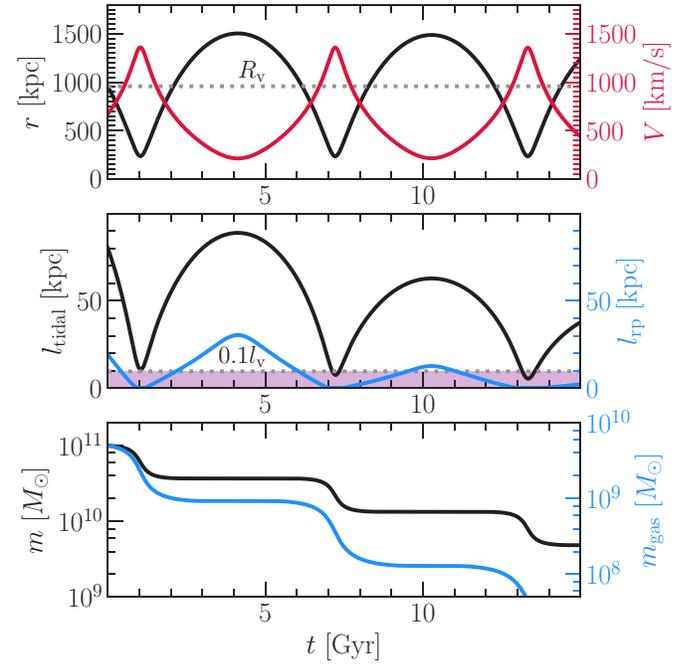} 
\caption{ Comparison of ram pressure stripping and the tidal stripping, for a UDG-sized satellite (of a virial mass $\mv=10^{11}\Msun$ and an NFW density profile with a concentration of $c=10$) orbiting a cluster/group-sized host (with $\Mv=10^{14}\Msun$ and $c=5$) along a typical cosmological orbit defined by $\xc=1$ and circularity $\eta=0.5$ (see text for definitions; $\xc=1$ means that the orbital energy is the same as that of a circular orbit with a radius of $\Rv$.)
We have assumed the gas distribution to be self-similar to the total mass profiles, scaled down by a gas fraction of 5\% of the total mass. 
\quad\quad
The {\it upper} panel shows the orbital radius ($r$) and velocity $V(r)$ as a function of time. 
The {\it middle} panel compares the tidal radius ($\lt$) and the ram pressure radius ($\lrp$, where ram pressure is equal to the gravitational restoring force per unit area).
The {\it bottom} panel compares the evolution of the total mass with that of the gas mass of the satellite (see text for details).
\quad\quad
The tidal radius is always larger than the ram pressure radius, which drops to zero around orbital pericentres, explaining the abrupt removal of cold gas that we have seen in the simulation result shown in \fig{Qhistories1}.
In the toy model, the gas mass decreases by an order of magnitude at pericentres while the subhalo mass decreases by $\sim2/3$, in good qualitative agreement with what we have seen in the simulations.
}
\label{fig:RamPressureRadius} 
\end{figure}

\smallskip 
We compare the importance of tides to that of ram pressure in stripping the cold gas of satellites by comparing their radii of influence.\footnote{In this section, we introduce several radii, which we summarize in Table 1.}
The radius of influence for tides within a satellite, i.e., the tidal radius ($\lt$), is usually defined as the distance to the satellite centre (along the line connecting the satellite and the host) where the self-gravity force toward the satellite centre is balanced by the tidal force from the host halo. 
We adopt the following expression of the tidal radius (e.g., \citealt{king62,zb03}),
\be\label{eq:lt}
\frac{m(\lt)}{M(r)} = \left[ 2 - \mu(r) + \frac{\vert V_{\rm t} \vert^2}{V_{\rm c}(r)^2} \right] \frac{\lt^3}{r^3}
\ee
where $m(l)$ and $M(r)$ are the mass profiles of the satellite and host, respectively; 
$\mu(r) = \rmd\ln M/\rmd\ln r$ at $r$ is the local slope of the host mass profile; 
$V_{\rm c}(r)=G M(r)/r$ is the circular velocity at $r$;
and $V_{\rm t}  = |\hat{\bm{r}} \times \bm{V}|$ is the instantaneous tangential velocity.  
While the first two terms inside the parentheses represent the gravitational tidal force, the third term represents the differential centrifugal force across the satellite due to its rotation about the halo centre. 
Note that near the halo characteristic radius, where the halo profile is close to an isothermal sphere, $\mu \sim 1$, if the tangential velocity at pericentre is comparable to the circular velocity, and the factor in parentheses is about 2. 
It becomes close to unity inside an inner halo cusp of $\mu \sim 2$, and it may vanish inside an inner core of $\mu = 3$ \citep{dekel03}.

\smallskip 
Similarly, one can define a ram pressure radius, $\lrp$, as the satellite-centric distance where the self-gravitational restoring force per unit area is equal to the ram pressure exerted by the gas of the host system on that of satellite. 
That is, $\lrp$ is the solution to the equation
\be\label{eq:lrp}
P_{\rm grav} (\lrp) = P_{\rm ram}(r),
\ee
where the restoring pressure $P_{\rm grav}$ is given by (e.g., \citealt{zinger18})
\be
P_{\rm grav}(l) = \frac{Gm(l)\rho_{\rm sg}(l)}{l},
\ee
with $\rho_{\rm sg}(l)$ being the gas density profile of the satellite galaxy; and the ram pressure $P_{\rm ram}$ is given by
\be
P_{\rm ram}(r) = \rho_{\rm hg}(r)V(r)^2
\ee
with $\rho_{\rm hg}(r)$ being the gas profile of the host, and $V(r)$ the velocity of the satellite with respect to the host. 

\smallskip 
To keep things representative for a UDG-sized object falling into a cluster/group-sized host halo, we consider a satellite with $\mv=10^{11}\Msun$ orbiting a host halo of $\Mv=10^{14}\Msun$, and assume that their density profiles, $\rho_{\rm sat}(l)$ and $\rho_{\rm host}(r)$, follow NFW profiles with $\cNFW=10$ and $5$, respectively.\footnote{we have verified that assuming a cored profile of the form of \equ{dekel} for the satellite yields results that are qualitatively similar.}
We assume the gas distributions to be self-similar to the total density profiles, and scaled down by the gas fractions of the satellite ($\fgs\equiv \mg/\mv$) and the host ($\fgh\equiv \Mg/\Mv$), respectively: 
\be
\rho_{\rm sg} = \fgs \rho_{\rm sat}(l),
\ee
\be
\rho_{\rm hg} = \fgh \rho_{\rm host}(r).
\ee
We adopt $\fgs=\fgh=0.05$ as the fiducial values, after having experimented with $\fgs$ and $\fgh$ ranging from 0.01 to 0.17, respectively, to confirm that the results shown in the following hold firmly.

\smallskip 
Both $\lt$ and $\lrp$ will vary along the orbit. 
Following a common convention, we specify an orbit with two parameters: first, an orbital energy proxy, $\xc\equiv \rc(E)/\Rv$, which is the radius of the circular orbit corresponding to the orbital energy $E$ in units of the virial radius of the host halo; second, the orbital circularity, $\eta\equiv j/\jc(E)$, which is the ratio between the specific orbital angular momemtum and the angular momentum of a circular orbit of energy $E$.  
We consider an orbit with $\xc=1$ and $\eta=0.5$ that is commonly found for satellites in cosmological simulations (e.g., \citealt{vdb17}).

\smallskip 
Crudely speaking, the {\it gas} outside the ram pressure radius tries to escape the satellite, so does {\it all} the mass outside the tidal radius.
But the affected mass will not be stripped off abruptly, but gradually over some timescale.
We assume the stripping timescale to be the local dynamical time at the host-centric radius $r$: 
\be\label{eq:tstrip}
\tau_{\rm strip}(r) = \tdyn(r) = \sqrt{ \frac{3\pi}{16G\rhobar(r)} },
\ee
where $\rhobar(r)$ is the average density of the host halo inside radius $r$. 
The instantaneous mass loss rate is therefore given by
\be\label{eq:massloss}
\frac{\rmd m}{\rmd t} = \frac{ m(>\lt) }{\tau_{\rm strip}(r)}
\ee
for the total mass\footnote{An alternative assumption of the stripping timescale that is commonly used in the literature is $\tau_{\rm strip}(r)=t_{\rm orb}(r)/\alpha$, where $t_{\rm orb}=2\pi/\Omega$, with $\Omega=V_{\rm t}/r$ the instantaneous angular speed of the satellite, and $\alpha$ is a free factor (e.g., \citealt{zb03}). 
We find that our assumption of $\tau_{\rm strip}(r)=t_{\rm dyn}(r)$ is almost equivalent to $\tau_{\rm strip}(r)=t_{\rm orb}(r)/\alpha$ with $\alpha=3$, a value similar to those found by matching the mass-loss rate to simulation results, e.g., $\alpha=3.5$ as reported by \cite{zentner05} and \cite{vdb18a}, $\alpha=2.5$ by \cite{pullen14}.}, and 
\be
\frac{\rmd \mg}{\rmd t} = \frac{ \mg[>\min(\lt,\lrp)] }{\tau_{\rm strip}(r)}
\ee
for the gas mass.

\smallskip 
We integrate the orbit starting from the initial virial-crossing and present the evolution of the radii of influence and the masses in \fig{RamPressureRadius}. 
We learn that for the assumed density profiles and orbit, the ram pressure radius is always smaller than the tidal radius, confirming that ram pressure is more important than tides in removing the gas content of the satellite.
Considering the inner $0.1\lv$ as the extent of the cold gas and the stars of satellite galaxy, we can see that the tidal radius dips only briefly into the baryonic extent, consistent with what we have seen in the simulations that the stellar mass is only marginally stripped for most of the surviving satellites. 
In contrast, the ram-pressure radius drops well below $0.1\lv$, especially near pericentres.
This explains the abrupt removal of gas at orbital pericentres that we have seen in the simulations. 
At each pericentre, the subhalo mass decreases by $\sim2/3$, while the gas mass decreases by almost an order of magnitude, also consistent with what we have seen in the simulations presented in \figs{Qhistories1}-\ref{fig:Qhistories2}.
We conclude that ram pressure stripping is the dominant factor in quenching satellites, including UDGs. 

\subsection{Tidal stripping and tidal heating}
\label{sec:heating}

\smallskip
We showed in \figs{Qhistories1} examples of normal dwarf galaxies turning into UDGs at orbital pericentres, where the half-stellar mass radii increase by $\sim50$ per cent in a couple of host dynamical times.
The size growth is associated with significant subhalo mass loss, a spike in the kinetic energy of the stars, and a brief deviation from virial equilibrium. 
Based on these phenomena, we argued that tidal heating operates actively along with tidal stripping during the transition from normal dwarfs to UDGs.

\smallskip
However, it is not trivial to quantify the contribution of tidal heating in puffing up the satellites, because heating and stripping occur simultaneously, with most of the kinetic energy injected into the satellite at the pericentre deposited to the particles on the outskirts of the satellite that will be stripped, and thus not contributing to the expansion of the remaining bound system. 
Trying to consolidate the relevance of tidal heating in puffing up satellite galaxies, we present here an analytic estimate of the kinetic energy $\DE$ injected to the {\it bound} part of a satellite over a full orbit. 
In particular, we compare $\DE$ with the initial binding energy\footnote{We define the binding energy $\Eb$ as the energy required to disassemble a system, so $\Eb$ is a {\it positive} number, equal to the absolute value of the sum of the total internal kinetic energy and the total potential energy of the system. }$\Eb$ of the satellite -- if $\DE/\Eb\ll1$, then tidal heating is irrelevant as a mechanism for puffing up satellites, while if $\DE/\Eb$ is a large fraction of unity, tidal heating is a viable mechanism of creating UDGs.

\smallskip
The ratio $\DE/\Eb$ depends on orbital parameters, so we will experiment with different orbital configurations.
We start by considering a fiducial orbit, with $\xc=1$ and $\eta=0.5$, and then generalize.

\subsubsection{Tidal heating energy}
\label{sec:gho99}

\begin{figure*}
\includegraphics[width=0.75\textwidth]{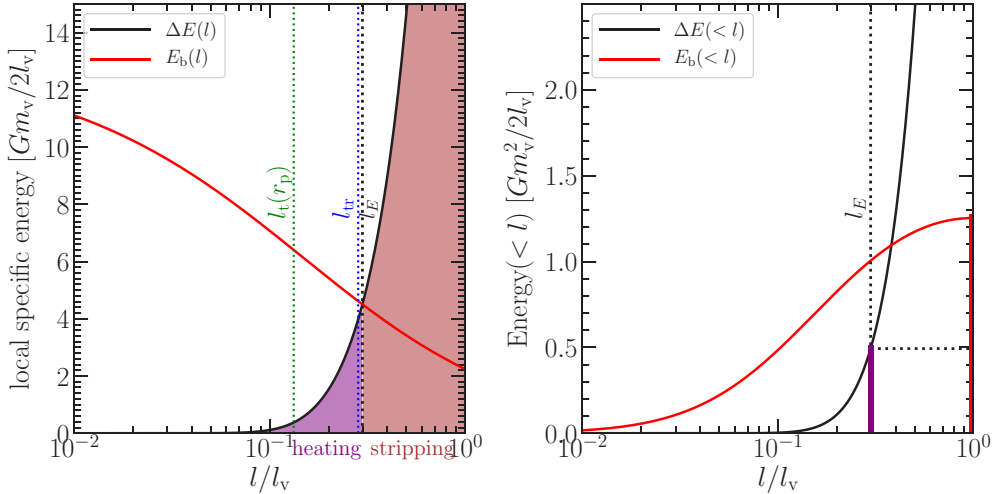} 
\caption{ 
Tidal heating of an NFW subhalo with virial mass $\mv=10^{11}\Msun$ and concentration $c=10$ during a full orbit of energy $\xc=1$ and circularity $\eta=0.5$, about a NFW host halo of virial mass $\Mv=10^{14}\Msun$ and concentration $c=5$.
{\it Left}: the specific heating energy $\DE(l)$ acquired in a full orbit as a function of the satellite-centric radius $l$ (black), compared with specific binding energy $\Eb(l)$ (red).
\quad
{\it Right}: total tidal heating acquired over a full orbit (black) compared with the total  binding energy (red), inside satellite-centric radius $l$.
\quad
The black, vertical dotted line marks the radius $\lE$ where the the local specific heating energy is equal to the local specific binding energy -- it is an estimate of the effective truncation radius over a full orbit. 
Tidal stripping provides an orthogonal estimate of the effective truncation radius, $\ltr$ (see text), indicated by the blue vertical dotted line, which agrees very well with $\lE$. 
The green vertical dotted line marks the instantaneous tidal radius at the pericentre, $\lt(\rp)$.
\quad
The radius $\lE$ divides the regime for stripping ($l>\lE$) and the regime where puffing up is expected.
Most of the tidal energy goes into the outer part that will be stripped, but the tidal energy inside the bound radius $\lE$ is still significant, amounting up to $\sim$40 per cent of the binding energy (as can be seen by comparing the purple and red bars in the right-hand panel).
Between $\lt(\rp)$ and $\lE$ is where tidal puffing up is optimal. 
}
\label{fig:lE} 
\end{figure*}

\smallskip
We follow the prescription of \cite{gho99} to estimate the tidal heating energy along an eccentric orbit, from one apocentre to the next, in a spherically symmetric host potential. 
We show that the average energy boost of a satellite particle at radius $l$ is given by (see Appendix \ref{sec:heating_app} for the derivation)
\be \label{eq:DE}
\DE(l) =  \frac{1}{6\lambda^2} \vv^2  \frac{l^2}{\lv^2}  \chi A(l),
\ee 
where 
\be
\lambda = \eta \xc^{1/2} [M(\xc\Rv)/\Mv]^{1/2}
\ee
is a dimensionless measurement of the orbital angular momentum; $\vv$ is the virial velocity of the satellite; the factor $\chi$ contains the information of the orbit; and the factor $A(l)$ that is less than unity corrects for the particles not in the impulsive regime. 

\smallskip
Neglecting the structural change that occurs over one orbit, one can calculate the total heating energy injected to the part of the satellite within a radius $l$:
\be \label{eq:DEtot}
\DE(<l) = 4\pi \int_0^{l} \DE(l^\prime)\rho(l^\prime)l^{\prime2}\rmd l^\prime,
\ee 
where $\DE(l)$ is given by \equ{DE} and $\rho(l)$ is the satellite's density profile.

\subsubsection{An effective tidal truncation radius: $\lE$}
\label{sec:lE}

\begin{table}
\caption{Definitions of radii used in this work.}
\begin{tabular}{llr} 
\hline
\hline
    & Definition & Equation \\
\hline
& {\it radii with respect to the host centre} &\\
$r$      & host-centric radius    & --     \\
$\rp$      & orbital peri-centre distance to host centre    & --     \\
$\Rv$          & host virial radius        & --      \\
\hline
& {\it radii with respect to the satellite centre} & \\
$l$          & satellite-centric radius        & --       \\
$\lv$          & satellite virial radius        & --       \\
$\lt$         & tidal radius         & \equ{lt}      \\
$\lrp$          & ram-pressure stripping radius         & \equ{lrp}      \\
$\lt(\rp)$          & tidal radius at orbital pericentre $\rp$  & --   \\
$\lE$          & tidal truncation radius after a full orbit    & \equ{lE}      \\
&  (based on tidal heating energy)  & \\
$\ltr$          & tidal truncation radius after a full orbit       & \equ{ltr}      \\
&  (based on tidal stripping of mass)  & \\
\hline
\end{tabular}
\label{table:radii}
\end{table}

\smallskip
We showed an estimate of the {\it instantaneous} tidal radius in the upper panel of \fig{RamPressureRadius}. 
Here we estimate the effective tidal truncation radius of a satellite {\it over a full orbit}.
There are multiple ways to do so.
From the perspective of energy balance, one can define the truncation radius ($\lE$) to be where the heating energy accumulated over a full orbit is equal to the local binding energy, i.e., the solution of 
\be \label{eq:lE}
\DE(l)=\Eb(l).
\ee

\smallskip
As before, we consider a satellite with an NFW density profile, a virial mass $\mv=10^{11}\Msun$, and a concentration parameter $c=10$, orbiting a cluster-sized host, also following an NFW profile with $\Mv=10^{14}\Msun$ and $c_{\rm h}=5$.
We compute the local specific heating energy $\DE(l)$ using \equ{DE} and the formalism detailed in Appendix \ref{sec:heating_app}, and the local specific binding energy $\Eb(l)$ using the energy identities of NFW profiles listed in Appendix \ref{sec:NFW},  and solve \equ{lE} for $\lE$.

\smallskip
As shown in the left-hand panel of \fig{lE}, we find $\lE$ to be $\simeq0.3\lv$ for our fiducial orbit (with $\xc=1$ and $\eta=0.5$). 
The tidal energy $\DE(l)$ increases with the satellite-centric radius $l$, gradually at $l<\lE$, and steeply at $l>\lE$. 
The radius $\lE$ divides two regimes: the particles at $l>\lE$ will be stripped, while the particles at $l<\lE$ will remain bound, with the tidal energy gain used to puff up the system.

\smallskip
In \fig{lE} we also mark the position of the instantaneous tidal radius at pericentre, $\lt(\rp)$, obtained using \equ{lt}.
The truncation radius $\lE$ is larger than the tidal radius at pericentre.
This is because tidal stripping is continuous over a timescale $\tau_{\rm strip}$ rather than abrupt.
As can be seen, the {\it regime of effective tidal heating} is approximately between the instant peri-centre tidal radius $\lt(\rp)$ and the truncation radius $\lE$.

\smallskip
One can alternatively estimate the truncation radius of a satellite over a full orbit from the perspective of tidal stripping.
Integrating the mass loss following \equ{massloss} over a full orbit\footnote{Again, we have assumed that the satellite evolves self-similarly along the orbit, i.e., the mass decreases but the concentration remains constant.}, we obtain the escaped mass $\Dm$, and define another truncation radius $\ltr$, as the solution to the equation
\be\label{eq:ltr}
m(\ltr) = \mv - \Dm,
\ee
where $m(l)$ is the initial mass profile of the satellite. 
It turns out that $\ltr$ is similar to the $\lE$ derived from the perspective of energy balance -- for our particular setup, $\ltr\simeq\lE$, and we have verified that $\ltr\simeq\lE$ is valid for a wide range of orbital parameters. 
Therefore, we adopt $\lE$ (or $\ltr$) as a robust estimate of the effective truncation radius of a satellite completing a full orbit. 

\subsubsection{Tidal heating of the bound part of the satellite}
\label{sec:DEE}

\smallskip
We compute the {\it total} heating energy injected to the part of the satellite inside radius $l$, $\DE(<l)$, using \equ{DEtot} and Appendix B.
The right-hand panel of \fig{lE} compares $\DE(<l)$ with the total binding energy $\Eb(<l)$ for the same setup. 
We can see that the total heating out to the virial radius $\lv$ is several times higher than the total binding energy. 
But this does not mean that the satellite will be disrupted in one orbit, since most of the tidal energy goes into the stripping regime at $l>\lE$. 
We can see that the part of tidal heating energy that can be utilized for puffing up the satellite is $\DE(<\lE)$, which in this case is $\sim40$ per cent of the binding energy,  i.e., a significant fraction that can change the internal structure of the satellite. 

\begin{figure*}
\includegraphics[width=\textwidth]{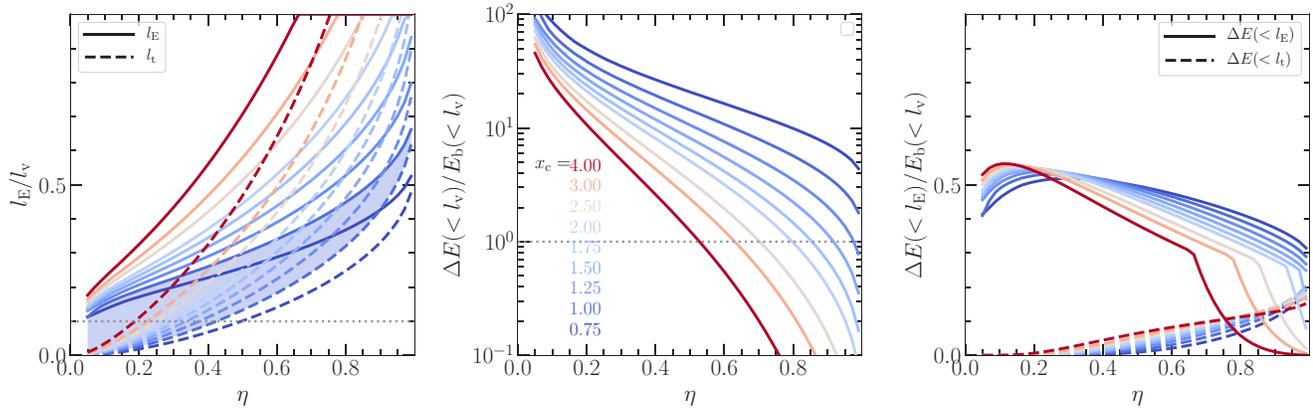} 
\caption{ 
Tidal heating of an NFW subhalo with virial mass $\mv=10^{11}\Msun$ and concentration $c=10$ during a full orbit about a host halo of virial mass $\Mv=10^{14}\Msun$ and concentration $c=5$, as a function of the orbit circularity $\eta$ (circular: $\eta=1$; radial: $\eta\to0$), and as a function of orbital energies (blue: low-energy; red: high-energy). 
\quad
{\it Left}: effective tidal truncation radius $\lE$ (where the local heating energy during a full orbit, $E(l)$, is equal to the local binding energy, $\Eb(l)$), in units of the original virial radius of the satellite. 
Overplotted in dashed lines are the tidal radii at pericentre, $\lt(\rp)$. 
The regime where tidal heating is effective is between $\lt(\rp)$ and $\lE$, and it varies with orbital circularity and energy -- the shaded blue band highlights the effective heating regime for $\xc=1$, i.e., an orbit with energy equal to the circular orbit of radius $\Rv$. (A typical cosmological orbit has $\xc\sim1$ and $\eta\sim0.5$.)
Assuming the stellar component of the satellite extends out to $0.1\lv$ (dotted line), for the stars to overlap with the efficient heating regime, the orbit needs to be highly eccentric ($\eta\la0.4$).
\quad
{\it Middle}: total tidal heating inside the virial radius $\lv$, in units of the total binding energy of the halo. 
Note that even for high-energy orbits, the total heating energy can be comparable or significantly larger than the binding energy. 
\quad
{\it Right}: Similar to the middle panel, but for the heating energy within the {\it bound} part of the satellite, $\DE(<\lE)$. 
The ratio between the purple bar and the red bar in the right-hand panel of \fig{lE} is a special case of what is plotted here.   
For most of the parameter space, $\DE(<\lE)$ amounts to $\sim30-55\%$ of the total binding energy $\Eb(<\lv)$.
The sharp decrease for the circular (high-$\eta$), high-energy (red) orbits is due to $\lE$ being larger than $\lv$ and thus capped at $\lv$.
}
\label{fig:DeltaE2} 
\end{figure*}

\smallskip
We repeat the above analysis for a range of orbital configurations, with orbital energy $\xc=0.75,1,1.25,...,4$ and orbital circularity $\eta=0.05,0.06,...,0.99$.
That is, we cover orbits ranging from nearly radial ($\eta=0.05$) to almost circular ($\eta=0.99$) with energies in the range that can be found for satellites in a cosmological simulation (e.g. \citealt{vdb17}). 
\Fig{DeltaE2} presents the truncation radius $\lE$, and the ratio of heating energy to binding energy $\DE/\Eb$, as functions of orbital circularity $\eta$ and energy $\xc$.
From the left-hand panel of \Fig{DeltaE2} one can read off the regime of effective tidal heating, i.e., between $\lt(\rp)$ and $\lE$, as a function of orbital configuration.
We have highlighted the case for $\xc=1$, which is most common in cosmological simulations, and it is clear that the effective heating regime overlaps with the baryonic range ($l<0.1\lv$) of the satellite galaxy for eccentric orbits ($\eta\la0.4$), while for $\eta>0.5$, the baryonic part of the satellites is not directly affected by tidal heating. 

%

\smallskip
The middle panel of \Fig{DeltaE2} shows the ratio of the tidal energy that goes into the whole satellite $\DE(<\lv)$ and the total binding energy $\Eb(<\lv)$, as a function of $\eta$ and $\xc$. 
For most of the parameter space, the tidal energy surpasses the binding energy by far -- though, again, most of the energy goes into the outskirts of the satellites that will be stripped and thus cannot be utilized for puffing up the system. 
The right-hand panel is similar to the middle one, but shows the heating energy that goes into the {\it bound} part of the satellite $\DE(<\lE)$.
We can see that, for most of the configurations, the heating energy amounts to $30$-$55$ per cent of the total binding energy $\Eb(<\lv)$.

\smallskip
We conclude that tidal heating is a feasible mechanism for puffing up satellite galaxies, but is only efficient for the satellites on sufficiently eccentric orbits and having stellar components extending to the regime between $\lt(\rp)$ and $\lE$. 

\subsection{Fate and origins of UDGs}
\label{sec:fractions} 

\smallskip
Since we have learnt that group UDGs can form in two pathways, as field UDGs that survive the group environmental effects and as less-diffuse dwarfs that are puffed up by tidal heating, it would be interesting to evaluate the relative contribution of each path to the population of group UDGs. 

{\it What is the fate of a field UDG when it falls into a cluster?}
We find in our group simulation that among the galaxies that entered the host halo as UDGs,
about 20 per cent survive (as UDGs) till $z=0$.  About 20 per cent manage to coalesce with the central galaxy, while about 60 percent are disrupted before they penetrate to the inner $0.15\Rv$ radius. 
Along the way to their current positions, the surviving UDGs are somewhat puffed up further by tides and become quiescent due to ram-pressure stripping. 
We do not see evidence in our simulations for field UDGs being more susceptible to tidal disruption than dwarfs that are less diffuse at infall. 
In fact, a similar 60 per cent of all the satellites that ever existed in the group have been disrupted. 
We caution that artificial disruption of satellite haloes may still be prevalent in modern simulations \citep{vdb18b}, and therefore refrain from over-interpreting the result. 

\smallskip
{\it What is the origin of a cluster UDG?}
Among all the satellites that have ever existed in the group and undergone a UDG stage, 20 per cent survive to the present day.
Among the surviving UDGs, 50 per cent originate from field UDGs (formed by SNe feedback), and the other half were normal galaxies that were puffed up by tides as satellites, i.e., the contributions from tidal puffing up and from the survival of field UDGs are comparable.  

\section{Conclusion}
\label{sec:conclusion}

\smallskip
In this study, we use the NIHAO simulations of field galaxies and a simulation of a galaxy group in an attempt to understand the origin of ultra-diffuse galaxies in the field and in a dense environment.
We use an analytic toy model to interpret the simulation results for group UDGs. 

\begin{itemize}
\smallskip
\item
We have extended the work of \cite{dicintio17} and shown that the field UDGs that lie in a characteristic narrow halo mass range, $\Mv = 10^{10.5\pm0.6}\Msun$, tend to be triaxial and prolate, far from rotating, exponential discs, but their S\'ersic indices are near unity. 
Their dark-matter density profile exhibits a flat density core dominating the regime within the stellar effective radius and is well described by the \cite{dekel17} function with $\aD=-1.5$ and $\cD=40$. 

\smallskip
\item
We find group UDGs have many properties in common with the field UDGs, especially having stellar masses and effective radii almost independent of the distance from the host-halo centre (For field galaxies in the group simulation, the host-centric distance is the actual distance from the group centre, which is $>\Rv$; and for the field galaxies in the NIHAO simulations, they are selected to be $\gg\Rv$ for any neighbouring massive halo). 
Satellite galaxies that survive generally lose only a small fraction of their stellar mass, explaining the null trend of satellite stellar mass with the host-centric distance.
The null trend of UDG size with host-centric distance is the outcome of two competing effects: on one hand, satellites grow in size (due to tidal heating, which is more efficient for satellites on lower-energy and more eccentric orbits); on the other hand, galaxies on the outskirts of the host system are accreted later, and therefore are larger due to progenitor bias. 

\smallskip
\item
We find a colour/sSFR gradient of group UDGs with distance from the host-halo centre, as observed.
Given the mild stellar mass evolution and the significant loss of gas mass at pericentres, we infer that it is ram pressure, rather than tides, that removes the gas from group UDGs when they are near orbital pericentres and quenches star formation. 

\smallskip
\item
We have identified two equally important origins of group UDGs. 
Satellite galaxies that were already UDGs at infall can survive the dense environment.
In addition, more compact field galaxies can get puffed up and become UDGs near orbital pericentres.
The size expansion is accompanied by energetics indicative of impulsive tidal heating. 
The expansion of the bound component of a satellite is associated with an increase of kinetic energy near pericentre, comparable to the tidal energy, followed by a relaxation to virial equilibrium at a larger size during the subsequent few dynamical times. 

\smallskip
\item 
We use simple analytic models to compare the roles of ram pressure and tides in stripping the cold gas from satellites, and to evaluate the importance of tidal heating in puffing up satellites.
We define the radius of influence of ram pressure as the satellite-centric distance where the self-gravitational restoring force per unit area is equal to the ram pressure, and show that the ram pressure radius is always smaller than the tidal radius throughout a typical orbit, confirming that the gas stripping is dominated by ram pressure.
Analytic calculations also indicate that tidal heating is a feasible mechanism for making UDGs from normal dwarfs on eccentric orbits.
In particular, the tidal energy deposited into the bound part of a satellite can amount to $\sim 50$ per cent of the total binding energy, in one orbit.
However, there are three regimes within the satellite -- the inner regime within the instantaneous tidal radius at pericentre, $\lt(\rp)$, where the tides are of little effect; the outer regime beyond an effective truncation radius, $\lE$, where the tides cause stripping; and an intermediate regime where the shells puff up but remain bound.
The inner regime and the intermediate regime will mix during the revirialization of the system.
For the stellar component of the satellite to fall in this intermediate regime of efficient tidal puffing up, a satellite has to be on a fairly eccentric orbit, with a circularity of $j/\jc(E)\la0.4$.
\end{itemize}

\section*{Acknowledgments}

We acknowledge stimulating discussions with Frank van den Bosch, Adi Nusser, Shany Danieli, Yuval Birnboim, Elad Zinger, and Nicolas Cornuault.  
This work was partly supported by the grants ISF 124/12, I-CORE Program of the PBC/ISF 1829/12, BSF 2014-273, PICS 2015-18, and NSF AST-1405962.
FJ is partly supported by the PBC Fellowship.
AJR was supported by NSF grant AST-1616710, and as a Research Corporation for Science Advancement Cottrell Scholar.
ADC acknowledges financial support from a Marie-Sk\l{}odowska-Curie Individual Fellowship grant, H2020-MSCA-IF-2016 Grant agreement 748213 DIGESTIVO.
The simulations were performed on the High Performance Computing resources at New York University Abu Dhabi; on the THEO cluster of the Max Planck Institute f${\rm \ddot{u}}$r Astronomie and on the HYDRA clusters at the Rechenzentrum in Garching. 

\bibliographystyle{mn2e}
\bibliography{udg}

\begin{thebibliography}{82}
\expandafter\ifx\csname natexlab\endcsname\relax\def\natexlab#1{#1}\fi

\bibitem[{Alabi {et~al}\mbox{.}(2018)Alabi, {Ferr{\'e}-Mateu, Anna},
  {Romanowsky, Aaron J}, {Brodie, Jean}, {Forbes, Duncan A}, {Wasserman,
  Asher}, {Bellstedt, Sabine}, {Martin-Navarro, Ignacio}, {Pandya, Viraj},
  {Stone, Maria B}, \& {Okabe, Nobuhiro}}]{alabi18}
Alabi A. {et~al.}, 2018, \mnras, 479, 3308

\bibitem[{Allgood {et~al}\mbox{.}(2006)Allgood, {Flores, Ricardo A}, {Primack,
  Joel R}, {Kravtsov, Andrey V}, {Wechsler, Risa H}, {Faltenbacher, Andreas},
  \& {Bullock, James S}}]{allgood06}
Allgood B., {Flores, Ricardo A}, {Primack, Joel R}, {Kravtsov, Andrey V},
  {Wechsler, Risa H}, {Faltenbacher, Andreas}, {Bullock, James S}, 2006,
  \mnras, 367, 1781

\bibitem[{Amorisco(2018)}]{amorisco18mass}
Amorisco N.~C., 2018, \mnras: Letters, 475, L116

\bibitem[{Amorisco \& Loeb(2016)}]{al16}
Amorisco N.~C., Loeb A., 2016, \mnras: Letters, 459, L51

\bibitem[{Amorisco {et~al}\mbox{.}(2018)Amorisco, Monachesi, Agnello, \&
  White}]{amorisco18gc}
Amorisco N.~C., Monachesi A., Agnello A., White S. D.~M., 2018, \mnras, 475,
  4235

\bibitem[{Beasley {et~al}\mbox{.}(2016)Beasley, Romanowsky, Pota, Navarro,
  Martinez~Delgado, Neyer, \& Deich}]{beasley16}
Beasley M.~A., Romanowsky A.~J., Pota V., Navarro I.~M., Martinez~Delgado D.,
  Neyer F., Deich A.~L., 2016, \apjl, 819, L20

\bibitem[{Behroozi {et~al}\mbox{.}(2014)Behroozi, Wechsler, Lu, Hahn, Busha,
  Klypin, \& Primack}]{behroozi14}
Behroozi P.~S., Wechsler R.~H., Lu Y., Hahn O., Busha M.~T., Klypin A., Primack
  J.~R., 2014, \apj, 787, 156

\bibitem[{Bose {et~al}\mbox{.}(2018)Bose, Frenk, Jenkins, Fattahi, Gomez,
  Grand, Marinacci, Navarro, Oman, Pakmor, Schaye, Simpson, \&
  Springel}]{bose18}
Bose S. {et~al.}, 2018, arXiv.org, arXiv:1810.03635

\bibitem[{Boylan-Kolchin, Ma \& Quataert(2008)Boylan-Kolchin, Ma, \&
  Quataert}]{bk08}
Boylan-Kolchin M., Ma C.-P., Quataert E., 2008, \mnras, 383, 93

\bibitem[{{Bullock} {et~al}\mbox{.}(2001){Bullock}, {Dekel}, {Kolatt},
  {Kravtsov}, {Klypin}, {Porciani}, \& {Primack}}]{bullock01}
{Bullock} J.~S., {Dekel} A., {Kolatt} T.~S., {Kravtsov} A.~V., {Klypin} A.~A.,
  {Porciani} C., {Primack} J.~R., 2001, \apj, 555, 240

\bibitem[{Burkert(2017)}]{burkert17}
Burkert A., 2017, \apj, 838, 93

\bibitem[{Carleton {et~al}\mbox{.}(2018)Carleton, {Errani, Raphael}, {Cooper,
  Michael}, {Kaplinghat, Manoj}, \& {Pe{\~n}arrubia, Jorge}}]{carleton18}
Carleton T., {Errani, Raphael}, {Cooper, Michael}, {Kaplinghat, Manoj},
  {Pe{\~n}arrubia, Jorge}, 2018, arXiv.org, arXiv:1805.06896

\bibitem[{Ceverino, Primack \& Dekel(2015)Ceverino, Primack, \&
  Dekel}]{ceverino15}
Ceverino D., Primack J., Dekel A., 2015, \mnras, 453, 408

\bibitem[{Chan {et~al}\mbox{.}(2018)Chan, Kere{\v s}, Wetzel, Hopkins,
  Faucher-Gigu{\`e}re, El-Badry, Garrison-Kimmel, \& Boylan-Kolchin}]{chan18}
Chan T.~K., Kere{\v s} D., Wetzel A., Hopkins P.~F., Faucher-Gigu{\`e}re C.~A.,
  El-Badry K., Garrison-Kimmel S., Boylan-Kolchin M., 2018, \mnras, 478, 906

\bibitem[{Danieli \& van Dokkum(2018)}]{danieli18}
Danieli S., van Dokkum P., 2018, arXiv.org, arXiv:1811.01962

\bibitem[{Dekel, Devor \& Hetzroni(2003)Dekel, Devor, \& Hetzroni}]{dekel03}
Dekel A., Devor J., Hetzroni G., 2003, \mnras, 341, 326

\bibitem[{Dekel {et~al}\mbox{.}(2017)Dekel, Ishai, Dutton, \& Maccio}]{dekel17}
Dekel A., Ishai G., Dutton A.~A., Maccio A.~V., 2017, \mnras, 468, 1005

\bibitem[{Di~Cintio {et~al}\mbox{.}(2017)Di~Cintio, Brook, Dutton, Macci{\`o},
  Obreja, \& Dekel}]{dicintio17}
Di~Cintio A., Brook C.~B., Dutton A.~A., Macci{\`o} A.~V., Obreja A., Dekel A.,
  2017, \mnras: Letters, 466, L1

\bibitem[{Di~Cintio {et~al}\mbox{.}(2014{\natexlab{a}})Di~Cintio, {Brook, C B},
  {Dutton, A A}, {Macci{\`o}, A V}, {Stinson, G S}, \& {Knebe,
  A}}]{dicintio14a}
Di~Cintio A., {Brook, C B}, {Dutton, A A}, {Macci{\`o}, A V}, {Stinson, G S},
  {Knebe, A}, 2014{\natexlab{a}}, \mnras, 441, 2986

\bibitem[{Di~Cintio {et~al}\mbox{.}(2014{\natexlab{b}})Di~Cintio, {Brook, C B},
  {Macci{\`o}, A V}, {Stinson, G S}, {Knebe, A}, {Dutton, A A}, \& {Wadsley,
  J}}]{dicintio14b}
Di~Cintio A., {Brook, C B}, {Macci{\`o}, A V}, {Stinson, G S}, {Knebe, A},
  {Dutton, A A}, {Wadsley, J}, 2014{\natexlab{b}}, \mnras, 437, 415

\bibitem[{Dutton \& Macci{\`o}(2014)}]{dm14}
Dutton A.~A., Macci{\`o} A.~V., 2014, \mnras, 441, 3359

\bibitem[{Dutton {et~al}\mbox{.}(2016)Dutton, Macci{\`o}, Dekel, Wang, Stinson,
  Obreja, Di~Cintio, Brook, Buck, \& Kang}]{dutton16}
Dutton A.~A. {et~al.}, 2016, \mnras, 461, 2658

\bibitem[{Dutton {et~al}\mbox{.}(2015)Dutton, Macci{\`o}, Stinson, Gutcke,
  Penzo, \& Buck}]{dutton15}
Dutton A.~A., Macci{\`o} A.~V., Stinson G.~S., Gutcke T.~A., Penzo C., Buck T.,
  2015, \mnras, 453, 2447

\bibitem[{Einasto(1965)}]{einasto}
Einasto J., 1965, Trudy Astrofizicheskogo Instituta Alma-Ata, 5, 87

\bibitem[{El-Badry {et~al}\mbox{.}(2016)El-Badry, Wetzel, Geha, Hopkins,
  Kere{\v s}, Chan, \& Faucher-Gigu{\`e}re}]{elbadry16}
El-Badry K., Wetzel A., Geha M., Hopkins P.~F., Kere{\v s} D., Chan T.~K.,
  Faucher-Gigu{\`e}re C.-A., 2016, \apj, 820, 131

\bibitem[{Errani, Pe{\~n}arrubia \& Walker(2018)Errani, Pe{\~n}arrubia, \&
  Walker}]{errani18}
Errani R., Pe{\~n}arrubia J., Walker M.~G., 2018, \mnras, 481, 5073

\bibitem[{Ferr{\'e}-Mateu {et~al}\mbox{.}(2018)Ferr{\'e}-Mateu, Alabi, Forbes,
  Romanowsky, Brodie, Pandya, Mart{\'\i}n-Navarro, Bellstedt, Wasserman, Stone,
  \& Okabe}]{fm18}
Ferr{\'e}-Mateu A. {et~al.}, 2018, \mnras, 479, 4891

\bibitem[{Gnedin, Hernquist \& Ostriker(1999)Gnedin, Hernquist, \&
  Ostriker}]{gho99}
Gnedin O.~Y., Hernquist L., Ostriker J.~P., 1999, \apj, 514, 109

\bibitem[{Gnedin \& Ostriker(1999)}]{go99}
Gnedin O.~Y., Ostriker J.~P., 1999, \apj, 513, 626

\bibitem[{Greco {et~al}\mbox{.}(2018)Greco, Greene, Strauss, Macarthur,
  Flowers, Goulding, Huang, Kim, Komiyama, Leauthaud, Leisman, Lupton,
  Sif{\'o}n, \& Wang}]{greco18}
Greco J.~P. {et~al.}, 2018, \apj, 857, 104

\bibitem[{Gu {et~al}\mbox{.}(2018)Gu, Conroy, Law, van Dokkum, Yan, Wake,
  Bundy, Merritt, Abraham, Zhang, Bershady, Bizyaev, Brinkmann, Drory,
  Grabowski, Masters, Pan, Parejko, Weijmans, \& Zhang}]{gu18}
Gu M. {et~al.}, 2018, \apj, 859, 37

\bibitem[{Harris, Blakeslee \& Harris(2017)Harris, Blakeslee, \&
  Harris}]{harris17}
Harris W.~E., Blakeslee J.~P., Harris G. L.~H., 2017, \apj, 836, 67

\bibitem[{Janssens {et~al}\mbox{.}(2017)Janssens, Abraham, Brodie, Forbes,
  Romanowsky, \& van Dokkum}]{janssens17}
Janssens S., Abraham R., Brodie J., Forbes D., Romanowsky A.~J., van Dokkum P.,
  2017, \apjl, 839, L17

\bibitem[{King(1962)}]{king62}
King I., 1962, \aj, 67, 471

\bibitem[{Knollmann \& Knebe(2009)}]{ahf}
Knollmann S.~R., Knebe A., 2009, \apjs, 182, 608

\bibitem[{Koda {et~al}\mbox{.}(2015)Koda, Yagi, Yamanoi, \& Komiyama}]{koda15}
Koda J., Yagi M., Yamanoi H., Komiyama Y., 2015, \apjl, 807, L2

\bibitem[{Komatsu {et~al}\mbox{.}(2011)Komatsu, Smith, Dunkley, Bennett, Gold,
  Hinshaw, Jarosik, Larson, Nolta, Page, Spergel, Halpern, Hill, Kogut, Limon,
  Meyer, Odegard, Tucker, Weiland, Wollack, \& Wright}]{wmap7}
Komatsu E. {et~al.}, 2011, \apjs, 192, 18

\bibitem[{Lange {et~al}\mbox{.}(2015)Lange, {Driver, Simon P}, {Robotham, Aaron
  S G}, {Kelvin, Lee S}, {Graham, Alister W}, {Alpaslan, Mehmet}, {Andrews,
  Stephen K}, {Baldry, Ivan K}, {Bamford, Steven}, {Bland-Hawthorn, Joss},
  {Brough, Sarah}, {Cluver, Michelle E}, {Conselice, Christopher J}, {Davies,
  Luke J M}, {Haeussler, Boris}, {Konstantopoulos, Iraklis S}, {Loveday, Jon},
  {Moffett, Amanda J}, {Norberg, Peder}, {Phillipps, Steven}, {Taylor, Edward
  N}, {L{\'o}pez-S{\'a}nchez, {\'A}ngel R}, \& {Wilkins, Stephen M}}]{lange15}
Lange R. {et~al.}, 2015, \mnras, 447, 2603

\bibitem[{Leisman {et~al}\mbox{.}(2017)Leisman, Haynes, Janowiecki, Hallenbeck,
  J{\'o}zsa, Giovanelli, Adams, Bernal~Neira, Cannon, Janesh, Rhode, \&
  Salzer}]{leisman17}
Leisman L. {et~al.}, 2017, \apj, 842, 133

\bibitem[{Lelli, McGaugh \& Schombert(2016)Lelli, McGaugh, \&
  Schombert}]{lelli16}
Lelli F., McGaugh S.~S., Schombert J.~M., 2016, \aj, 152, 157

\bibitem[{Lim {et~al}\mbox{.}(2018)Lim, Peng, Cote, Sales, den Brok, Blakeslee,
  \& Guhathakurta}]{lim18}
Lim S., Peng E.~W., Cote P., Sales L.~V., den Brok M., Blakeslee J.~P.,
  Guhathakurta P., 2018, arXiv.org, arXiv:1806.05425

\bibitem[{Ludlow {et~al}\mbox{.}(2013)Ludlow, Navarro, Boylan-Kolchin, Bett,
  Angulo, Li, White, Frenk, \& Springel}]{ludlow13}
Ludlow A.~D. {et~al.}, 2013, \mnras, 432, 1103

\bibitem[{Lupi, Volonteri \& Silk(2017)Lupi, Volonteri, \& Silk}]{lupi17}
Lupi A., Volonteri M., Silk J., 2017, \mnras, 470, 1673

\bibitem[{Marigo {et~al}\mbox{.}(2008)Marigo, Girardi, Bressan, Groenewegen,
  Silva, \& Granato}]{marigo08ssp}
Marigo P., Girardi L., Bressan A., Groenewegen M. A.~T., Silva L., Granato
  G.~L., 2008, \aap, 482, 883

\bibitem[{Mart{\'i}nez-Delgado {et~al}\mbox{.}(2016)Mart{\'i}nez-Delgado,
  L{\"a}sker, Sharina, Toloba, Fliri, Beaton, Valls-Gabaud, Karachentsev,
  Chonis, Grebel, Forbes, Romanowsky, Gallego-Laborda, Teuwen,
  G{\'o}mez-Flechoso, Wang, Guhathakurta, Kaisin, \& Ho}]{md16}
Mart{\'i}nez-Delgado D. {et~al.}, 2016, \apj, 151, 96

\bibitem[{Merritt {et~al}\mbox{.}(2016)Merritt, van Dokkum, Danieli, Abraham,
  Zhang, Karachentsev, \& Makarova}]{merritt16}
Merritt A., van Dokkum P., Danieli S., Abraham R., Zhang J., Karachentsev
  I.~D., Makarova L.~N., 2016, \apj, 833, 168

\bibitem[{Mowla {et~al}\mbox{.}(2017)Mowla, van Dokkum, Merritt, Abraham, Yagi,
  \& Koda}]{mowla17}
Mowla L., van Dokkum P., Merritt A., Abraham R., Yagi M., Koda J., 2017, \apj,
  851, 27

\bibitem[{Navarro, Frenk \& White(1997)Navarro, Frenk, \& White}]{nfw}
Navarro J.~F., Frenk C.~S., White S. D.~M., 1997, \apj, 490, 493

\bibitem[{Norris {et~al}\mbox{.}(2014)Norris, Kannappan, Forbes, Romanowsky,
  Brodie, Faifer, Huxor, Maraston, Moffett, Penny, Pota, Smith-Castelli,
  Strader, Bradley, Eckert, Fohring, McBride, Stark, \& Vaduvescu}]{norris14}
Norris M.~A. {et~al.}, 2014, \mnras, 443, 1151

\bibitem[{Pandya {et~al}\mbox{.}(2018)Pandya, Romanowsky, Laine, Brodie,
  Johnson, Glaccum, Villaume, Cuillandre, Gwyn, Krick, Lasker,
  Mart{\'\i}n-Navarro, Martinez-Delgado, \& van Dokkum}]{pandya18}
Pandya V. {et~al.}, 2018, \apj, 858, 29

\bibitem[{Papastergis, Adams \& Romanowsky(2017)Papastergis, Adams, \&
  Romanowsky}]{papastergis17}
Papastergis E., Adams E. A.~K., Romanowsky A.~J., 2017, \aap, 601, L10

\bibitem[{{Planck Collaboration}(2016)}]{planck15}
{Planck Collaboration}, 2016, A\&A, 594, A13

\bibitem[{Pontzen \& Governato(2012)}]{pg12}
Pontzen A., Governato F., 2012, \mnras, 421, 3464

\bibitem[{{Pontzen} {et~al}\mbox{.}(2013){Pontzen}, {Ro{\v s}kar}, {Stinson},
  {Woods}, {Reed}, {Coles}, \& {Quinn}}]{pynbody}
{Pontzen} A., {Ro{\v s}kar} R., {Stinson} G.~S., {Woods} R., {Reed} D.~M.,
  {Coles} J., {Quinn} T.~R., 2013, {pynbody: Astrophysics Simulation Analysis
  for Python}. Astrophysics Source Code Library, ascl:1305.002

\bibitem[{Pullen, Benson \& Moustakas(2014)Pullen, Benson, \&
  Moustakas}]{pullen14}
Pullen A.~R., Benson A.~J., Moustakas L.~A., 2014, \apj, 792, 24

\bibitem[{Rom{\'a}n \& Trujillo(2017)}]{rt17}
Rom{\'a}n J., Trujillo I., 2017, \mnras, 468, 4039

\bibitem[{Rong {et~al}\mbox{.}(2017)Rong, Guo, Gao, Liao, Xie, Puzia, Sun, \&
  Pan}]{rong17}
Rong Y., Guo Q., Gao L., Liao S., Xie L., Puzia T.~H., Sun S., Pan J., 2017,
  \mnras, 470, 4231

\bibitem[{Ruiz-Lara {et~al}\mbox{.}(2018)Ruiz-Lara, Beasley,
  Falc{\'o}n-Barroso, Rom{\'a}n, Pinna, Brook, Cintio, Mart{\'\i}n-Navarro,
  Trujillo, \& Vazdekis}]{rl18}
Ruiz-Lara T. {et~al.}, 2018, \mnras, 478, 2034

\bibitem[{{Shen}, {Wadsley} \& {Stinson}(2010){Shen}, {Wadsley}, \&
  {Stinson}}]{shen10}
{Shen} S., {Wadsley} J., {Stinson} G., 2010, \mnras, 407, 1581

\bibitem[{Shi {et~al}\mbox{.}(2017)Shi, Zheng, Zhao, Pan, Li, Zou, Zhou, Guo,
  An, \& Li}]{shi17}
Shi D.~D. {et~al.}, 2017, \apj, 846, 26

\bibitem[{Starkman {et~al}\mbox{.}(2018)Starkman, Lelli, McGaugh, \&
  Schombert}]{starkman18}
Starkman N., Lelli F., McGaugh S., Schombert J., 2018, \mnras, 480, 2292

\bibitem[{{Stinson} {et~al}\mbox{.}(2006){Stinson}, {Seth}, {Katz}, {Wadsley},
  {Governato}, \& {Quinn}}]{stinson06}
{Stinson} G., {Seth} A., {Katz} N., {Wadsley} J., {Governato} F., {Quinn} T.,
  2006, \mnras, 373, 1074

\bibitem[{Stinson {et~al}\mbox{.}(2013)Stinson, Bovy, Rix, Brook, Ro{\v s}kar,
  Dalcanton, Macci{\`o}, Wadsley, Couchman, \& Quinn}]{stinson13}
Stinson G.~S. {et~al.}, 2013, \mnras, 436, 625

\bibitem[{Taffoni {et~al}\mbox{.}(2003)Taffoni, {Mayer, Lucio}, {Colpi,
  Monica}, \& {Governato, Fabio}}]{taffoni03}
Taffoni G., {Mayer, Lucio}, {Colpi, Monica}, {Governato, Fabio}, 2003, \mnras,
  341, 434

\bibitem[{Tollet {et~al}\mbox{.}(2016)Tollet, {Macci{\`o}, Andrea V}, {Dutton,
  Aaron A}, {Stinson, Greg S}, {Wang, Liang}, {Penzo, Camilla}, {Gutcke, Thales
  A}, {Buck, Tobias}, {Kang, Xi}, {Brook, Chris}, {Di Cintio, Arianna},
  {Keller, Ben W}, \& {Wadsley, James}}]{tollet16}
Tollet E. {et~al.}, 2016, \mnras, 456, 3542

\bibitem[{Tomassetti {et~al}\mbox{.}(2016)Tomassetti, {Dekel, Avishai},
  {Mandelker, Nir}, {Ceverino, Daniel}, {Lapiner, Sharon}, {Faber, Sandra},
  {Kneller, Omer}, {Primack, Joel}, \& {Sai, Tanmayi}}]{tomassetti16}
Tomassetti M. {et~al.}, 2016, \mnras, 458, 4477

\bibitem[{van~den Bosch(2017)}]{vdb17}
van~den Bosch F.~C., 2017, \mnras, 468, 885

\bibitem[{van~den Bosch {et~al}\mbox{.}(2016)van~den Bosch, {Jiang, Fangzhou},
  {Campbell, Duncan}, \& {Behroozi, Peter}}]{vdb16}
van~den Bosch F.~C., {Jiang, Fangzhou}, {Campbell, Duncan}, {Behroozi, Peter},
  2016, \mnras, 455, 158

\bibitem[{van~den Bosch \& Ogiya(2018)}]{vdb18b}
van~den Bosch F.~C., Ogiya G., 2018, \mnras, 475, 4066

\bibitem[{van~den Bosch {et~al}\mbox{.}(2018)van~den Bosch, Ogiya, Hahn, \&
  Burkert}]{vdb18a}
van~den Bosch F.~C., Ogiya G., Hahn O., Burkert A., 2018, \mnras, 474, 3043

\bibitem[{van~der Wel {et~al}\mbox{.}(2014)van~der Wel, Franx, van Dokkum,
  Skelton, Momcheva, Whitaker, Brammer, Bell, Rix, Wuyts, Ferguson, Holden,
  Barro, Koekemoer, Chang, McGrath, H{\"a}ussler, Dekel, Behroozi, Fumagalli,
  Leja, Lundgren, Maseda, Nelson, Wake, Patel, Labb{\'e}, Faber, Grogin, \&
  Kocevski}]{vdw14}
van~der Wel A. {et~al.}, 2014, \apj, 788, 28

\bibitem[{van Dokkum {et~al}\mbox{.}(2016)van Dokkum, Abraham, Brodie, Conroy,
  Danieli, Merritt, Mowla, Romanowsky, \& Zhang}]{vd16}
van Dokkum P. {et~al.}, 2016, \apjl, 828, L6

\bibitem[{van Dokkum {et~al}\mbox{.}(2017)van Dokkum, Abraham, Romanowsky,
  Brodie, Conroy, Danieli, Lokhorst, Merritt, Mowla, \& Zhang}]{vd17}
van Dokkum P. {et~al.}, 2017, \apjl, 844, L11

\bibitem[{van Dokkum {et~al}\mbox{.}(2015)van Dokkum, Abraham, Merritt, Zhang,
  Geha, \& Conroy}]{vd15a}
van Dokkum P.~G., Abraham R., Merritt A., Zhang J., Geha M., Conroy C., 2015,
  \apjl, 798, L45

\bibitem[{Wadsley, Keller \& Quinn(2017)Wadsley, Keller, \& Quinn}]{wadsley17}
Wadsley J.~W., Keller B.~W., Quinn T.~R., 2017, \mnras, 471, 2357

\bibitem[{Wadsley, Veeravalli \& Couchman(2008)Wadsley, Veeravalli, \&
  Couchman}]{wadsley08}
Wadsley J.~W., Veeravalli G., Couchman H. M.~P., 2008, \mnras, 387, 427

\bibitem[{Wang {et~al}\mbox{.}(2015)Wang, {Dutton, Aaron A}, {Stinson, Gregory
  S}, {Macci{\`o}, Andrea V}, {Penzo, Camilla}, {Kang, Xi}, {Keller, Ben W}, \&
  {Wadsley, James}}]{wang15}
Wang L., {Dutton, Aaron A}, {Stinson, Gregory S}, {Macci{\`o}, Andrea V},
  {Penzo, Camilla}, {Kang, Xi}, {Keller, Ben W}, {Wadsley, James}, 2015,
  \mnras, 454, 83

\bibitem[{Wolf {et~al}\mbox{.}(2010)Wolf, {Martinez, Gregory D}, {Bullock,
  James S}, {Kaplinghat, Manoj}, {Geha, Marla}, {Mu{\~n}oz, Ricardo R}, {Simon,
  Joshua D}, \& {Avedo, Frank F}}]{wolf10}
Wolf J., {Martinez, Gregory D}, {Bullock, James S}, {Kaplinghat, Manoj}, {Geha,
  Marla}, {Mu{\~n}oz, Ricardo R}, {Simon, Joshua D}, {Avedo, Frank F}, 2010,
  \mnras, 406, 1220

\bibitem[{Yagi {et~al}\mbox{.}(2016)Yagi, Koda, Komiyama, \& Yamanoi}]{yagi16}
Yagi M., Koda J., Komiyama Y., Yamanoi H., 2016, \apjs, 225, 11

\bibitem[{Zentner {et~al}\mbox{.}(2005)Zentner, Berlind, Bullock, Kravtsov, \&
  Wechsler}]{zentner05}
Zentner A.~R., Berlind A.~A., Bullock J.~S., Kravtsov A.~V., Wechsler R.~H.,
  2005, \apj, 505

\bibitem[{Zentner \& Bullock(2003)}]{zb03}
Zentner A.~R., Bullock J.~S., 2003, \apj, 598, 49

\bibitem[{Zinger {et~al}\mbox{.}(2018)Zinger, Dekel, Kravtsov, \&
  Nagai}]{zinger18}
Zinger E., Dekel A., Kravtsov A.~V., Nagai D., 2018, \mnras, 475, 3654

\end{thebibliography}

\appendix

\section{Analytic formalism for estimating impulsive heating energy}
\label{sec:heating_app}

\smallskip
In this appendix, we derive the expression for the impulsive heating energy as used in the main text \equ{DE},  following \cite{gho99}.

\smallskip
In the impulse approximation where the internal motions of the satellite particles are neglected, the change in velocity of a particle is given by 
\be \label{eq:Dv}
\Delta \mathbf{v} = \int_{-T_{\rm orb}/2}^{T_{\rm orb}/2} \mathbf{a}_{\rm tid} {\rmd} t,
\ee 
where $T_{\rm orb}$ is the radial period of the orbit, and $\mathbf{a}_{\rm tid}$ is the tidal acceleration, given by
\be
\mathbf{a}_{\rm tid}(r)=\frac{G\Mv}{r^3} \left[ (3\mu(r) -\mubar(r))(\hat{\mathbf{r}}\cdot\mathbf{r})\hat{\mathbf{r}} - \mu(r)\mathbf{r} \right],
\ee 
with 
 \be
\mu(r) \equiv \frac{M(r)}{\Mv}
\ee
denoting the normalized mass profile of the host, and 
\be
\mubar(r) \equiv \frac{\rmd \mu(r)}{\rmd \ln r}.
\ee

\smallskip
It is our freedom to choose the orbit to be in the $X$-$Y$ plane, and define the position angle $\theta$ to be $0$ at the pericentre.
Using the identity $\rmd t= r^2 /j \rmd \theta$, where $j$ is the specific orbital angular momentum, we can rewrite \equ{Dv} as
\be 
\Delta \mathbf{v} = \frac{1}{\lambda} \frac{\Vv}{\Rv} \left\{ (B_1-B_3)X, (B_2-B_3)Y, -B_3Z\right\}.
\ee
where $\lambda\equiv j/(\Rv\Vv)$ is a dimensionless angular momentum that can be expressed with the orbital parameters, $\xc$ and $\eta$, as
\begin{align}
\lambda & = \frac{j}{\Rv\Vv} \nonumber \\
& = \frac{j}{\jc(E)} \frac{\jc(E)}{[GM(\rc)\rc(E)]^{1/2}} \left[\frac{\rc(E)}{\Rv} \right] ^{1/2} \left[\frac{M(\rc)}{\Mv}\right]^{1/2} \nonumber \\
& = \eta \xc^{1/2} \mu(\xc\Rv)^{1/2};
\end{align}
and $B_i$ are integrals given by 
\begin{align}
B_1 & = \int_{-\thetam}^{\thetam}  \frac{3\mu(x) -\mubar(x)}{x} \cos^2(\theta) \rmd \theta, \label{eq:B1}\\
B_2 & = \int_{-\thetam}^{\thetam}  \frac{3\mu(x) -\mubar(x)}{x} \sin^2(\theta) \rmd \theta, \\
B_3 & = \int_{-\thetam}^{\thetam}  \frac{\mu(x)}{x} \rmd \theta,
\end{align}
with $x\equiv r/\Rv$ and $\thetam$ representing the position angle of the apocentre.

\smallskip
Computing the integrals $B_i$ requires the orbital radius as a function of position angle, $r(\theta)$, which we obtain by numerically integrating the orbit with a fifth-order Runge-Kutta method, and applying a cubic-spline fit to the part of the orbit between two apocentres. 
In practice, integrating from $\theta=-\pi/2$ to $\pi/2$ is accurate to percent level compared to using the actual apocentre position angel $\thetam$, because most of the heating occur near the pericentre. 

\smallskip
The average energy boost of a satellite particle at radius $l$ is given by
\be \label{eq:DEappendix}
\DE(l) = \frac{1}{2}\langle \Delta\mathbf{v}\cdot\Delta \mathbf{v} \rangle(l)=  \frac{1}{6\lambda^2} \vv^2  \frac{l^2}{\lv^2}  \chi A(l),
\ee 
where the factor $\chi$ contains the orbital information, given by
\be
\chi \equiv (B_1 - B_3)^2 + (B_2-B_3)^2 + B_3^2;
\ee 
and following \cite{go99}, we have added a correction factor $A(l)$($<1$), to account for the particles not in the impulsive regime. 
The {\it adiabatic correction} factor is empirically given by 
\be \label{eq:AC}
A(l) = [1+\omega(l)^2 \tau ]^{-\gamma}, 
\ee 
where $\omega(l)$ is the orbital frequency; $\tau$ is the duration of the encounter; and $\gamma$ is an empirical exponent.
Following \cite{vdb18a}, we use $\omega(l)=\sigmar(l)/l$, with $\sigmar(l)$ being the radial velocity dispersion; and $\tau=\rp/\Vp$, i.e., the paricentre distance divided by the speed of the satellite with respect to the host at the pericentre. 
We adopt $\gamma=5/2$, which is appropriate for fast encounters with $\tau<\tdyn$ \citep{go99}.
Note that, apart from the adiabatic correction, $\DE(l)\propto l^2$. 

\section{Energy structure of NFW profiles }
\label{sec:NFW}

\smallskip
The following identities are used for computing the effective tidal truncation radius $\lE$, and for evaluating the tidal heating energy $\DE(<l)$ as well as the binding energy $\Eb(<l)$ of an NFW profile. 

\smallskip
For an NFW profile, the normalized mass profile is 
\be \label{eq:M_NFW}
\mu(l) = \frac{f(y)}{f(c)},
\ee
where $y\equiv c l/\lv$, and $f(y)=\ln(1+y)-y/(1+y)$.

\smallskip
When computing the adiabatic correction, we use an approximate expression of $\sigmar(l)$ provided by \cite{zb03}.

\smallskip
The specific binding energy at radius $l$ is given by
\be
\Eb(l)=|T(l)+\phi(l)|, 
\ee
where $T(l)$ is the specific kinetic energy and $\phi(l)$ is the gravitational potential.
For analytic convenience, we assume that all the particles are on circular orbits, so that
\footnote{This gives a lower bound of the kinetic energy. Generally, $T(l)=(3-2\beta)\sigmar^2(l)/2$, where $\beta=1-\sigmat^2/\sigmar^2$ is the velocity anisotropy parameter.}
\be
T(l) = \frac{\vc^2(l)}{2},
\ee 
For an NFW profile, the potential is
\be 
\phi(l) = -\vv^2 \frac{c}{f(c)} \frac{\ln(1+y)}{y}.
\ee

\smallskip
For an NFW density profile, \equ{DEtot} can be written as
\be 
\DE(<l) =  \frac{ \chi }{6\lambda^2} \mv\vv^2 \frac{ 1}{c^2 f(c)}K(l),
\ee
where the factor $K$ is given by an integral
\be 
K(l) = \int_0^{x=cl/\lv} \frac{x^{\prime 3}A(x^\prime)}{(1+x^\prime)^2}\rmd x^\prime,
\ee
which, without the adiabatic correction ($A(x^\prime)$, defined in \equ{AC}), reduces to a simple expression, 
\be 
K(l) = \frac{x(x(x-3)-6)}{2(1+x)} + 3\ln(1+x).
\ee

\smallskip
The total binding energy is given by:
\be 
\Eb(<l) =  T(<l) + \Uin(<l) + \Uout(<l).
\ee
It is easy to show that, for an NFW profile,
\be 
T(<l) = \mv\vv^2 \frac{c}{2f(c)^2} I(x), 
\ee
is the total kinetic energy inside the radius $l$, where  
\be
I(x) = \frac{1}{2} - \frac{\ln(1+x)}{1+x} - \frac{1}{2(1+x)^2}
\ee
and $x=cl/\lv$.
Again, we have assumed that all particles are on circular orbits.
The potential energy contributed by the mass inside $l$ is given by
\be
\Uin(<l) = -\mv\vv^2 \frac{c}{2f(c)^2} 2I(x)=-2T(<l);
\ee
and the potential energy contributed by the mass outside $l$ is given by
\be
\Uout(<l) = -\mv\vv^2 \frac{c}{2f(c)^2}  f(x)\left[ \frac{1}{1+x} - \frac{1}{1+c} \right].
\ee

\label{lastpage}
\end{document}